\documentclass[a4paper,preprintnumbers,showpacs,onecolumn,superscriptaddress,nofootinbib,amsmath,amssymb,notitlepage]{revtex4-1}

\usepackage{enumitem}
\usepackage{times}
\usepackage{dcolumn}
\usepackage{mathrsfs}

\usepackage{comment}
\usepackage{hyperref}
\usepackage{amsmath}
\usepackage{mathtools}
\usepackage{bbm}
\usepackage{bm}
\usepackage{amsfonts}
\usepackage{amssymb}
\usepackage{mathrsfs}
\usepackage{wasysym}
\usepackage{graphicx}
\usepackage[]{subfigure}
\usepackage{filecontents}
\usepackage[dvipsnames]{xcolor}

\hypersetup{
    bookmarks=true,         
    pdftoolbar=true,        
    pdfmenubar=true,        
    pdffitwindow=true,     
    pdfstartview={FitH},     
    pdftitle={My title},     
    pdfauthor={author},      
    pdfsubject={Subject},    
    pdfcreator={Creator},    
    pdfproducer={Producer},  
    pdfkeywords={keyword1} 
    pdfnewwindow=true,       
    colorlinks=true ,        
    linkcolor=Blue  ,        
    citecolor=Blue,      
    filecolor=green,       
    urlcolor=Blue            
}

\newcommand{\arctanh}{\operatorname{arctanh}}

\newcommand{\ed}{\mathrm{d}}

\newcommand{\tL}{\tilde{\Lambda}}

\begin{document}

\title{Analytical study of particle geodesics around a scale-dependent 
de Sitter black hole}

\author{Mohsen Fathi}
\email{mohsen.fathi@usach.cl; mohsen.fathi@gmail.com}
\affiliation{Departamento de F\'{i}sica, Universidad de Santiago de Chile,
Avenida V\'{i}ctor Jara 3493, Estaci\'{o}n Central, 9170124, Santiago, Chile}

\begin{abstract}
We give a fully analytical description of radial and angular geodesics for
massive particles that travel in the spacetime provided by a $(3+1)$-dimensional
scale-dependent black hole in the cosmological background, for which, the
quantum corrections are assumed to be small. We show that the equations of
motion for radial orbits can be solved by means of Lauricella hypergeometric
functions with different numbers of variables. We then classify the angular
geodesics and argue that no planetary bound orbits are available. We calculate
the epicyclic frequencies of circular orbits at the potential's maximum and the
deflection angle of scattered particles is also calculated. Finally, we resolve
the raised Jacobi inversion problem for the angular motion by means of a
genus-2 Riemannian theta function, and the possible orbits are derived and
discussed.  \\

{\noindent{\textit{keywords}}: Black holes, time-like geodesics, scale-dependent gravity, cosmological constant}\\

\end{abstract}

\maketitle

\section{Introduction}

The reconciliation of geometry and gravity by the general theory of relativity is shown by the investigation of free-falling objects in the gravitational fields, where the curvature of spacetime plays the main role. In fact, the argument that planets and light do travel on geodesics was the main reason that general relativity could receive some popularity soon after its birth. In this regard, and after the proposition of the Schwarzschild solution \cite{1916SPAW189S}, the prediction and measurement of light deflection around the Sun, proven during the 1919 solar eclipse expedition \cite{1920RSPTA.220..291D}, and the accurate evaluation of the anomalous precession in the perihelion of Mercury \cite{RevModPhys.19.361}, can be named as the first two primary tests of general relativity. In fact, according to the non-linear nature of the partial differential equations that appear in the dynamics of moving particles in curved spacetimes, the above observational tests and the similar ones which are still in progress, have been based on the simplified results obtained from the approximate or numerical manipulations of the geodesic equations. On the other hand, it is of significant advantage to have in hand the analytical expressions. First, because they may serve as the touchstone for the numerical methods and approximations, and second, they can be used to make a complete systematic study of the parameter space, and hence, to make further predictions of the astrophysical observables. Accordingly, and since Hagihara's 1931 studies on the geodesics of particles in Schwarzschild spacetime \cite{1930JaJAG...8...67H}, which was then followed by
Darwin, Mielnik, and Pleba\'{n}ski \cite{noauthor_gravity_1959,noauthor_gravity_1961,Mielnik:1962}, 
efforts to find exact analytical solutions for the geodesic equations of massive and massless particles have been on the rise. In particular, the application of modular forms in solving the arising (hyper-)elliptic integrals in the study of geodesics has received considerable attention in the last two decades. These methods which are based on the theories of elliptic functions and modular forms, were studied by nominated nineteenth-century mathematicians such as Jacobi \cite{jacobi_2013}, Abel \cite{abel_2012}, Riemann \cite{Riemann:1857,Riemann+1866+161+172}, and Weierstrass \cite{Weierstrass+1854+289+306} (see also Ref. \cite{baker_abelian_1995} for a complete textbook review on these discoveries). Accordingly, numerous investigations have been devoted to the analysis of the time-like and null geodesics in static and stationary black hole spacetimes inferred from general relativity and its extensions, in which, the raised (hyper-)elliptic integrals are treated by means of hypergeometric, elliptic, and the Riemannian theta functions of the different genus (see for example Refs.  \cite{kraniotis_general_2002,kraniotis_compact_2003,kraniotis_precise_2004,kraniotis_frame_2005,cruz_geodesic_2005,kraniotis_periapsis_2007,hackmann_complete_2008,hackmann_geodesic_2008,hackmann_analytic_2009,hackmann_complete_2010,olivares_motion_2011,kraniotis_precise_2011,cruz_geodesic_2013,villanueva_photons_2013,kraniotis_gravitational_2014,soroushfar_analytical_2015,soroushfar_detailed_2016,hoseini_analytic_2016,hoseini_study_2017,fathi_motion_2020,fathi_classical_2020,fathi_gravitational_2021,gonzalez_null_2021,fathi_analytical_2021,kraniotis_gravitational_2021,fathi_study_2022,soroushfar_analytical_2022,battista_geodesic_2022,fathi_spherical_2023}). 

It is, however, important to mention that although general relativity has appeared successful in the course of the aforementioned astrophysical tests, this theory has not yet answered the long-lasting questions concerning the quantum nature of gravity. Hence, it is indispensable to search for a consistent theory of quantum gravity, which is one of the famous quests in modern theoretical physics.  In fact, in most cases, scientists try to take into account the scale-dependence of the gravitational action's couplings, once the quantum effects appear \cite{jacobson_thermodynamics_1995,connes_gravity_1996,connes_gravity_1996-1,rovelli_loop_1998,gambini_consistent_2005,ashtekar_gravity_2005,nicolini_noncommutative_2009,horava_quantum_2009,verlinde_origin_2011}. In this sense, the theories of gravity become scale-dependent (SD) at the quantum level. The SD theories of gravity have received notable attention during the last years, and in particular, their relevance to black hole spacetimes has been studied widely \cite{koch_scale_2016,rincon_scale-dependent_2017,rincon_quasinormal_2018,contreras_scale-dependent_2018,rincon_scale-dependent_2018,contreras_five-dimensional_2020,rincon_scale-dependent_2020,Panotopoulos:2021,rincon_four_2021}.

Also in this work, we consider a special $(3+1)$-dimensional (4D) static spherically symmetric SD spacetime for black holes in the cosmological background given in Ref. \cite{Panotopoulos:2021}. In the same interest as at the beginning of this section, we study and derive the exact analytical solutions to the equations of motion for massive particles moving in the exterior geometry of this black hole. The analysis requires a precise treatment of hyper-elliptic integrals with special properties, and we provide several methods for deriving the solutions. In particular, we exploit the Lauricella hypergeometric functions of different numbers of variables. Note that, the definite Lauricella functions have been used to calculate the period of planar and non-planar bound orbits in black hole spacetimes. In this study, for the first time, we present the indefinite Lauricella functions as the solutions to hyper-elliptic integrals that appear in the calculation of the radial geodesics and use them to simulate the possible orbits. The paper is organized as follows: In Sect. \ref{sec:theBH} we provide a brief introduction to the SD theory and introduce the black hole solution. This is followed by the derivation of the horizons and the causal structure of the spacetime. In Sect. \ref{sec:dynamics}, we construct the Lagrangian dynamics which is used to study the geodesics. In Sect. \ref{sec:radial} we begin our discussion, starting from the radial geodesics. The relevant effective potential and the corresponding types of orbits are derived and discussed in detail. In Sect. \ref{sec:angular}, we switch to the angular geodesics, which includes the analysis of the effective potential and the possible orbits. In this section, the scattering angle of deflected particles and the stability of circular orbits are also discussed. Within the paper, all kinds of orbits are plotted appropriately to demonstrate their properties. We conclude in Sect. \ref{sec:conclusion}. Throughout this work, we apply a geometrized unit system, in which $G=c=1$. Also wherever appears, prime denotes differentiation with respect to the $r$-coordinate.

\section{The SD black hole solution}\label{sec:theBH}

\begin{figure}
    \centering
    \includegraphics[width=8cm]{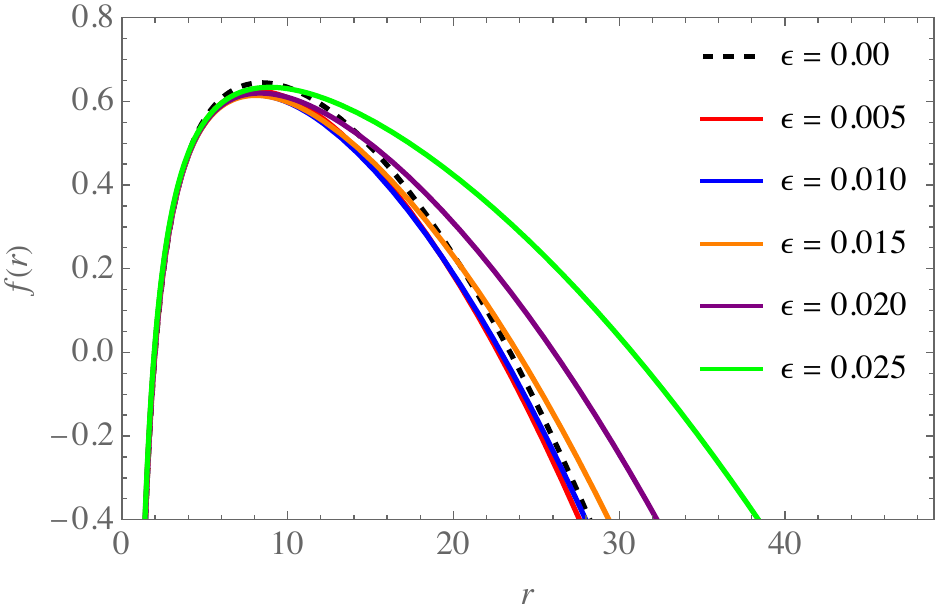}~(a)\qquad
     \includegraphics[width=8cm]{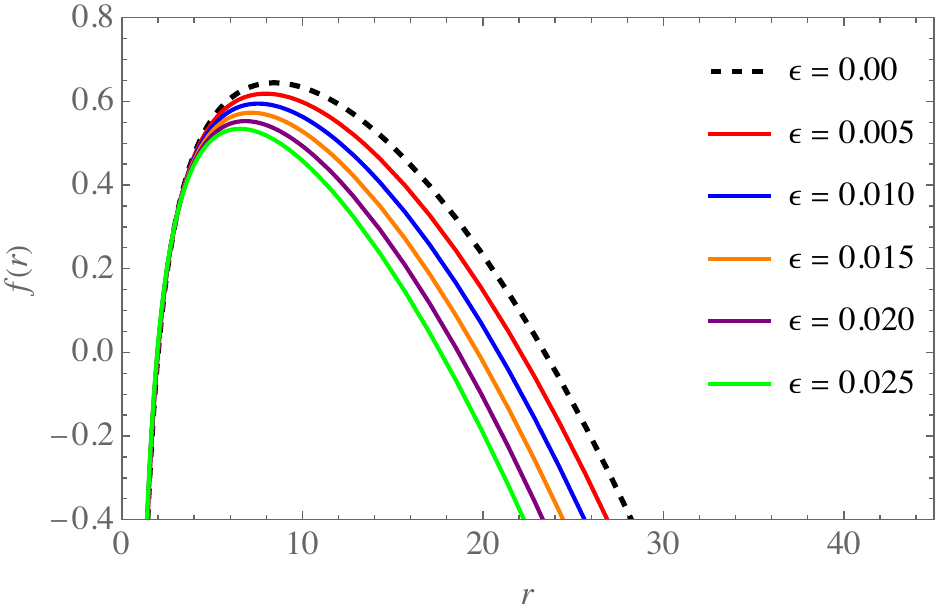}~(b)
    \caption{The radial profile of $f(r)$, plotted for different values of $\epsilon$ and $\tL=3\times 10^{-4}$. The diagrams correspond to the profiles of (a) the original lapse function \eqref{eq:lapse}, and (b) the first order estimation in Eq.~\eqref{eq:lapse1}. }
    \label{fig:f(r)}
\end{figure}

In the SD theories of gravity, the classical general relativistic solutions are extended by means of some SD coupling parameters that compensate for the quantum corrections. For the particular case which is of interest in this study, there are two coupling parameters that contribute to the construction of the theory; the \textit{running} cosmological constant $\Lambda_k$ and the running Newton's gravitational constant $G_k$, where $k(r)$ plays the role of an arbitrary re-normalization scale. This way, and by including the metric tensor $g_{\mu\nu}$ as the main ingredient, the Einstein field equations take the SD form \cite{Panotopoulos:2021}
\begin{equation}
G_{\mu\nu} + \Lambda_k g_{\mu\nu} = \kappa_k T_{\mu\nu}^{\mathrm{eff}}, 
    \label{eq:GR-eq0}
\end{equation}
in which $\kappa_k=8\pi G_k$, and the effective energy-momentum tensor is given by
\begin{equation}
{\kappa_k T_{\mu\nu}^{\mathrm{eff}} = \kappa_k T_{\mu\nu}^{\mathrm{M}}-\Delta t_{\mu\nu}},
    \label{eq:Teff}
\end{equation}
in terms of the matter $T_{\mu\nu}^{\mathrm{M}}$, and the $G$-varying
\begin{equation}
\Delta t_{\mu\nu} = G(r)\left(g_{\mu\nu}\square-\nabla_\mu\nabla_\nu\right)G^{-1}(r),
    \label{eq:G-varyingt}
\end{equation}
parts of the energy-momentum tensor, where $\square\equiv\nabla^\lambda\nabla_\lambda$. Now the null energy condition implies that $\mathcal{O}\left(k(r)\right)\rightarrow \mathcal{O}(r)$, and hence, only the radial variations are considered. This way, the static spacetime of the SD de Sitter (SDdS) black hole is found as
\begin{equation}\label{eq:metric}
    \ed s^2 = -f(r) \ed t^2+f^{-1}(r) \ed r^2 + r^2\ed\theta^2+r^2\sin^2\theta\ed\phi^2,
\end{equation}
in the usual Schwarzschild coordinates $x^\mu = (t,r,\theta,\phi)$, where the lapse function is given by \cite{Panotopoulos:2021}
\begin{equation}\label{eq:lapse}
    f(r) = 1-\frac{M\left[-2+3r\epsilon(1-2r\epsilon)\right]}{r}+\frac{r}{6}\left(-6\epsilon+9\epsilon r^2-2\Lambda r\right)\\
    +r^2\epsilon^2\left(1+6M\epsilon\right)\ln\left(1+\frac{1}{r\epsilon}\right),
\end{equation}
describing the exterior geometry of an object of mass $M$. Here, $\Lambda>0$ is the classical cosmological constant, and $\epsilon>0$ is the SD running parameter. Recently, this solution has been analyzed in Ref. \cite{ovgun_4d_2023} regarding the shadow and the deflection angle of light rays.  Note that $[\Lambda]=\mathrm{m}^{-2}$ and $[\epsilon]=\mathrm{m}^{-1}$. In this study, we consider small effects from the quantum corrections, in the sense that only up to the first order of the term $\epsilon r$ is taken into account. This way, we recast the lapse function \eqref{eq:lapse} as
\begin{equation}\label{eq:lapse1}
    f(r) = 1-\frac{2M}{r}+\left(3M-r\right)\epsilon-\frac{1}{3}\Lambda r^2.
\end{equation}
In Ref.~\cite{PhysRevD.103.104040}, this particular form has been used to study the analytical solutions for propagating null geodesics. As expected, for $\epsilon\rightarrow0$ the classical Schwarzschild-de Sitter spacetime is recovered. To facilitate the calculations, we do the transformation $r\rightarrow M r$, which is equivalent to letting $M=1$. We also let $\Lambda/3\rightarrow\tL$. The causal structure of this spacetime is determined by means of the solutions to the equation $f(r)=0$, which results in the three values
\begin{eqnarray}
&& r_1 = -\frac{4}{\tL}\sqrt{\frac{g_2}{3}}\cos\left(
\frac{1}{3}\arccos\left(
\frac{3g_3}{g_2}\sqrt{\frac{3}{g_2}}\right)-\frac{4\pi}{3}
\right)-\frac{\epsilon}{3\tL},\label{eq:r1} \\
&& r_2 =-\frac{4}{\tL}\sqrt{\frac{g_2}{3}}\cos\left(
\frac{1}{3}\arccos\left(
\frac{3g_3}{g_2}\sqrt{\frac{3}{g_2}}\right)-\frac{2\pi}{3}
\right)-\frac{\epsilon}{3\tL},\label{eq:r2}\\
&& r_{3}=-\frac{4}{\tL}\sqrt{\frac{g_2}{3}}\cos\left(
\frac{1}{3}\arccos\left(
\frac{3g_3}{g_2}\sqrt{\frac{3}{g_2}}\right)\right)-\frac{\epsilon}{3\tL},\label{eq:r3}
\end{eqnarray}
in which
\begin{subequations}\label{eq:g2g3}
\begin{align}
& g_2 =\frac{1}{12} \left(3 \tL+\epsilon ^2+9 \tL  \epsilon \right),\\
& g_3 =\frac{1}{432} \left(54 \tL ^2+2 \epsilon ^3+27 \tL  \epsilon ^2+9 \tL  \epsilon \right).
\end{align}
\end{subequations}
The discriminant of the equation $f(r)=0$ is of the form $\Delta=g_2^3-27g_3^2\approx\frac{\tL^3}{64}(1-27\tL)+\mathcal{O}(\epsilon^2)$, which is always positive for $\tL\ll1$. Hence, the radii in Eqs.~\eqref{eq:r1}--\eqref{eq:r3} are real-valued and it is straightforward to check that $r_1>r_2>0$ and $r_3<0$. Accordingly, the black hole has a cosmological horizon at $r_{++}=r_1$ and an event horizon at $r_+=r_2$. This way, the lapse function can be recast as
\begin{equation}
    f(r) = \frac{\tL}{r}\left(r_{++}-r\right)\left(r-r_+\right)\left(r-r_3\right).
    \label{eq:lapse_2}
\end{equation}
In Fig.~\ref{fig:f(r)}, the radial profile of the lapse function $f(r)$ has been plotted for some small values for the $\epsilon$-parameter.

\section{Lagrangian dynamics for motion of massive particles}\label{sec:dynamics}


The motion of massive particles in the spacetime provided by the line element \eqref{eq:metric}, can be described by the Lagrangian
\begin{eqnarray}
2\mathcal{L}&=&g_{\mu\nu}\dot x^\mu\dot x^\nu\nonumber\\
&=&-f(r)\dot t^2+\frac{\dot r^2}{f(r)}+r^2\left(\dot\theta^2+\sin^2\theta\dot\phi^2\right),
    \label{eq:Lagrangian}
\end{eqnarray}
in which $\dot{x}^\mu\equiv\ed x^\mu/\ed\tau$, where $\tau$ is the affine parameter of the geodesic curves. One can consider the conjugate momenta
\begin{equation}
\Pi_\mu = \frac{\partial\mathcal{L}}{\partial\dot{x}^\mu},
    \label{eq:conjMoment}
\end{equation}
which based on the Killing symmetries of the spacetime, introduces the two constants of motion
\begin{eqnarray}
&&\Pi_t\equiv-f(r)\dot t\equiv-E,\label{eq:E}\\
&&\Pi_\phi=r^2\dot\phi\equiv L,\label{eq:L}
    \label{eq:E,L}
\end{eqnarray}
with $E$ and $L$, termed respectively, as the energy and the angular momentum of the test particles\footnote{Note that, $E$ cannot be considered as the particles' energy since the spacetime is not asymptotically flat. It, however, presents a constant of motion which is fundamental in the categorization of the orbits, the same as energy.}. The time-like trajectories are distinguished by letting $2\mathcal{L}=-1$. This way, and by confining ourselves to the equatorial plane (i.e. $\theta=\pi/2$), the equations of motion are obtained as
\begin{eqnarray}
&& \dot r^2=E^2-V(r),\label{eq:dotr}\\
&& \left(\frac{\ed r}{\ed t}\right)^2 = \frac{f(r)^2}{E^2}\left[E^2-V(r)\right],\\
&& \left(\frac{\ed r}{\ed\phi}\right)^2=\frac{r^4}{L^2}\left[E^2-V(r)\right],\label{eq:drdphi}
\end{eqnarray}
in which 
\begin{equation}
V(r) = f(r)\left(1+\frac{L^2}{r^2}\right),
    \label{eq:V(r)}
\end{equation}
is the effective gravitational potential felt by the approaching particles. We begin our investigation by studying the radial trajectories.

\section{Radial motion}\label{sec:radial}

\begin{figure}
    \centering
    \includegraphics[width=9cm]{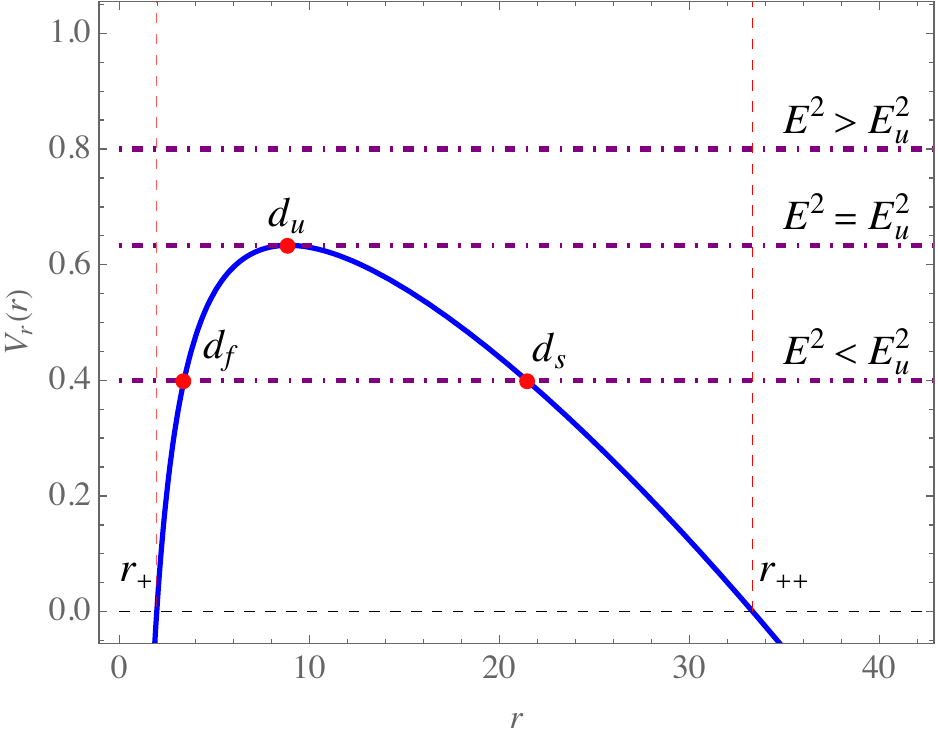}
    \caption{The radial effective potential, together with examples of the turning points, plotted for $\epsilon=0.02$ and $\tL=3\times 10^{-4}$.  In this case, the maximum radial distance for unstable orbits is $d_u=8.86$ for which $E^2=E^2_u=0.63$. The two turning points $d_s = 21.44$ and $d_f=3.40$ have been also shown, whose corresponding energy value is $E^2=0.4$. The turning point $d_s$ indicates the distance, at which $0<E^2<E_u^2$, and the frontal scattering of particles occurs.   
}
    \label{fig:Vr(r)}
\end{figure}

The study of infalling particles with zero angular momentum can have numerous advantages regarding the standard general relativistic tests. For example, the theoretical foundations of the so-called gravitational clock effect for falling observers in the gravitational fields are based on radial orbits, which is also related to the gravitational redshift-blueshift of light rays passing a black hole. Another example is the well-known \textit{frozen} infalling objects when they are observed by distant observers as they approach the black hole's event horizon. This is related to the difference between the perception of comoving and distant observers, as they observe infalling objects onto the black hole \cite{ryder_2009,zeldovich_stars_2014}. In this case, the effective potential takes the form $V_r(r)=f(r)$, whose radial profile has been shown in Fig.~\ref{fig:Vr(r)}.
The effective potential exhibits a maximum, so the motion becomes unstable where $V_r'(r)=0$, which results in the radial position
\begin{equation}
d_u = \frac{2}{\tL}\sqrt{\frac{\tilde{g}_2}{3}}\cos\left(\frac{1}{3}\arccos\left(
\frac{3\tilde{g}_3}{\tilde{g}_2}\sqrt{\frac{3}{\tilde{g}_2}}
\right)\right)-\frac{\epsilon}{6\tL},
    \label{eq:du}
\end{equation}
where
\begin{subequations}
\begin{align}
    &\tilde{g}_2 = \frac{\epsilon^2}{12},\\
    & \tilde{g}_3 = \frac{\tL^2}{4}\left(2-\frac{\epsilon^3}{54\tL^2}\right).
\end{align}
\label{eq:gt2,gt3}
\end{subequations}
The value in Eq.~\eqref{eq:du}, is the maximum distance for unstable orbits, which corresponds to the energy $E_u\equiv V_r(d_u)$. This way, one can categorize the radial orbits as follows: 
\begin{itemize}
    \item \textit{Frontal scattering of the first and second kinds (FSFK and FSSK)}: 
    For $0<E^2<E_u^2$, the orbits correspond to the FSFK when they encounter the turning point $d_s$ (for which $d_u<d_s<r_{++}$), or to the FSSK when they start from the turning point $d_f$ (for which $r_+<d_f<d_u$). In the case of the FSFK, the particles recede from the black hole after scattering, while for the FSSK, the particles fall inexorably onto the event horizon.
    
    \item \textit{Critical radial orbits}: In the case of $E^2=E_u^2$, depending on the initial distance of approach, the test particles encounter different fates. In this sense, when the particles approach from $d_i$ (for which $d_u<d_i<r_{++}$), they fall on the radius $d_u$, whereas when they come from $d_0$ (for which $r_+<d_0<d_u$), they are captured by the black hole. These two categories constitute the critical radial orbit of the first and second kinds (CROFK and CROSK). 
    
    \item \textit{Radial capture}: For $E^2>E_u^2$, the particles coming from a finite distance $d_j$ (for which $r_+<d_j<r_{++}$), will fall onto the event horizon.
\end{itemize}
In fact, by using the expression in Eq.~\eqref{eq:lapse_2}, the equations of motion for radial orbits can be rewritten as
\begin{eqnarray}
&& \dot r^2 = \frac{\mathfrak{p}_3(r)}{r},\label{eq:dotr_rad}\\
&& \left(\frac{\ed r}{\ed t}\right)^2=\frac{\tL^2(r_{++}-r)^2(r-r_+)^2(r-r_3)^2\mathfrak{p}_3(r)}{E^2r^3},\label{eq:drdt_rad}
\end{eqnarray}
in which
\begin{equation}
\mathfrak{p}_3(r) = \tL  r^3-  \left(r_3+r_++r_{++}\right)\tL r^2 + \left[E^2+  r_3 (r_++r_{++})\tL+  r_+ r_{++}\tL\right] r- r_3 r_+ r_{++}\tL.
    \label{eq:frakp(r)}
\end{equation}

\subsection{FSFK and FSSK}\label{subsec:FSFK,FSSK}

The characteristic polynomial \eqref{eq:frakp(r)} vanishes at the radial distances
\begin{eqnarray}
&& d_1 = \frac{4}{\tL}\sqrt{\frac{\bar{g}_2}{3}}\cos\left(\frac{1}{3}\arccos\left(
\frac{3\bar{g}_3}{\bar{g}_2}\sqrt{\frac{3}{\bar{g}_2}}
\right)-\frac{4\pi}{3}\right)+\frac{r_{++}+r_++r_3}{3},\label{eq:d1}\\
&& d_2 = \frac{4}{\tL}\sqrt{\frac{\bar{g}_2}{3}}\cos\left(\frac{1}{3}\arccos\left(
\frac{3\bar{g}_3}{\bar{g}_2}\sqrt{\frac{3}{\bar{g}_2}}
\right)-\frac{2\pi}{3}\right)+\frac{r_{++}+r_++r_3}{3},\label{eq:d2}\\
&& d_3 = \frac{4}{\tL}\sqrt{\frac{\bar{g}_2}{3}}\cos\left(\frac{1}{3}\arccos\left(
\frac{3\bar{g}_3}{\bar{g}_2}\sqrt{\frac{3}{\bar{g}_2}}
\right)\right)+\frac{r_{++}+r_++r_3}{3},\label{eq:d3}
\end{eqnarray}
where
\begin{subequations}
\begin{align}
    & \bar{g}_2 = \frac{\tL}{12}\left[
    \left(r_{++}+r_++r_3\right)^2\tL-3\left(E^2+r_+r_{++}\tL+r_3\left(r_++r_{++}\right)\tL\right)
    \right],\label{eq:gb2}\\
    & \bar{g}_3 = -\frac{\tL^2}{16}\left[
    \frac{1}{3}\left(r_{++}+r_++r_3\right)\left(E^2+r_+r_{++}\tL+r_3\left(r_++r_{++}\right)\tL\right)
    -r_3 r_+ r_{++}\tL-\frac{2}{27}\left(r_{++}+r_++r_3\right)^3\tL
    \right].
\end{align}
\end{subequations}
One can verify that $0<d_2<d_3$ and $d_1<0$. Hence, we can assign $d_f\equiv d_2$ and $d_s\equiv d_3$ at which, the frontal scatterings occur. This way, the characteristic polynomial can be recast as 
\begin{equation}
    \mathfrak{p}_3(r) = \tL \left(r-d_s\right)\left(r-d_f\right)\left(r-{d_1}\right).
    \label{eq:frakp(r)_1}
\end{equation}
By taking advantage of this simple form in the case of the FSFK at $r=d_s$, the equation of motion \eqref{eq:dotr_rad} leads to a degenerate hyper-elliptic integral which yields the solution (see appendix \ref{app:A})
\begin{equation}
\tau(r) = \frac{2 d_s}{\sqrt{\tL\ell^2}}\sqrt{1-\frac{d_s}{r}}F_D^{(3)}\left(
\frac{1}{2},b_1,b_2,1;\frac{3}{2};c_1,c_2,1-\frac{d_s}{r}
\right),
\label{eq:tau(r)_FSFK}    
\end{equation}
where $\ell^2=(d_s-d_f)(d_s-d_1)$, and $F_D^{(3)}$ is the incomplete 2-variable Lauricella hypergeometric function, which here can be given in terms of the one-dimensional Euler-type integral \cite{Exton:1976,Akerblom:2005}
\begin{equation}
\int_0^{1-d_s/r} z^{-\frac{1}{2}} (1-z)^{-1}\prod_{i=1}^{2}(1-c_i z)^{-b_i}\,\ed z = 2\sqrt{1-\frac{d_s}{r}}\, F_D^{(3)}\left(
\frac{1}{2},b_1,b_2,\frac{1}{2};\frac{3}{2};c_1,c_2,1-\frac{d_s}{r}
\right),
    \label{eq:FD3-tau(r)-FSFK}
\end{equation}
with $b_1=b_2=1/2$, and 
\begin{subequations}
\begin{align}
    & c_1 = -\frac{d_f}{d_s-d_f},\\
    & c_2 = -\frac{d_1}{d_s-d_1}.
\end{align}
\label{eq:ci-tau(r)-FSFK}
\end{subequations}
The solution given in Eq.~\eqref{eq:tau(r)_FSFK} relates to the perception of comoving observers with the radial geodesics in the course of the FSFK. {Note that, although expressing the solution in the form \eqref{eq:tau(r)_FSFK} is brief and aesthetically pleasant, nevertheless, the equation of motion \eqref{eq:dotr_rad} can still be solved in terms of ordinary elliptic integrals and Jacobi elliptic functions (see appendix \ref{app:B}).}
To the distant observers, the radial evolution of the coordinate time is obtained by solving the degenerate hyper-elliptic integral resulting from the equation of motion \eqref{eq:drdt_rad}, which yields (see Eq.~\eqref{eq:A5})
\begin{equation}
t(r) = -\frac{E}{2d_s^2\bar{\ell}^3\sqrt{\tL^3\bar{\bar{\ell}}^2}}\left(1-\frac{d_s}{r}\right)^2
F_D^{(6)}\left(2,1,1,1,\frac{1}{2},\frac{1}{2},\frac{1}{2};3;\bar{c}_1,\bar{c}_2,\bar{c}_3,\bar{c}_4,\bar{c}_5,1-\frac{d_s}{r}\right),
    \label{eq:t(r)_FSFK}
\end{equation}
in which $\bar{\ell}^3=-d_s^{-3}(r_{++}-d_s)(d_s-r_+)(d_s-r_3)$, $\bar{\bar{\ell}}^{2}=d_s^{-2}(d_s-d_f)(d_s-d_1)$, and
\begin{subequations}
\begin{align}
    & \bar{c}_1=-\frac{r_{++}}{d_s-r_{++}},\\
    & \bar{c}_2=-\frac{r_+}{d_s-r_{+}},\\
    & \bar{c}_3 = -\frac{r_3}{d_s-r_3},\\
    & \bar{c}_4=-\frac{d_f}{d_s-d_f},\\
    & \bar{c}_5=-\frac{d_1}{d_s-d_1}.
\end{align}
\label{eq:ci-t(r)-FSFK}
\end{subequations}
{Similar to the previous case, the equation of motion can be solved in terms of elliptic integrals, as explained in detail in appendix \ref{app:B}.} In Fig.~\ref{fig:FSFK,FSSK}, the radial profile of the time parameters have been plotted for the FSFK and FSSK, based on the solutions in Eqs.~\eqref{eq:tau(r)_FSFK} and \eqref{eq:t(r)_FSFK}.
\begin{figure}
    \centering
    \includegraphics[width=7cm]{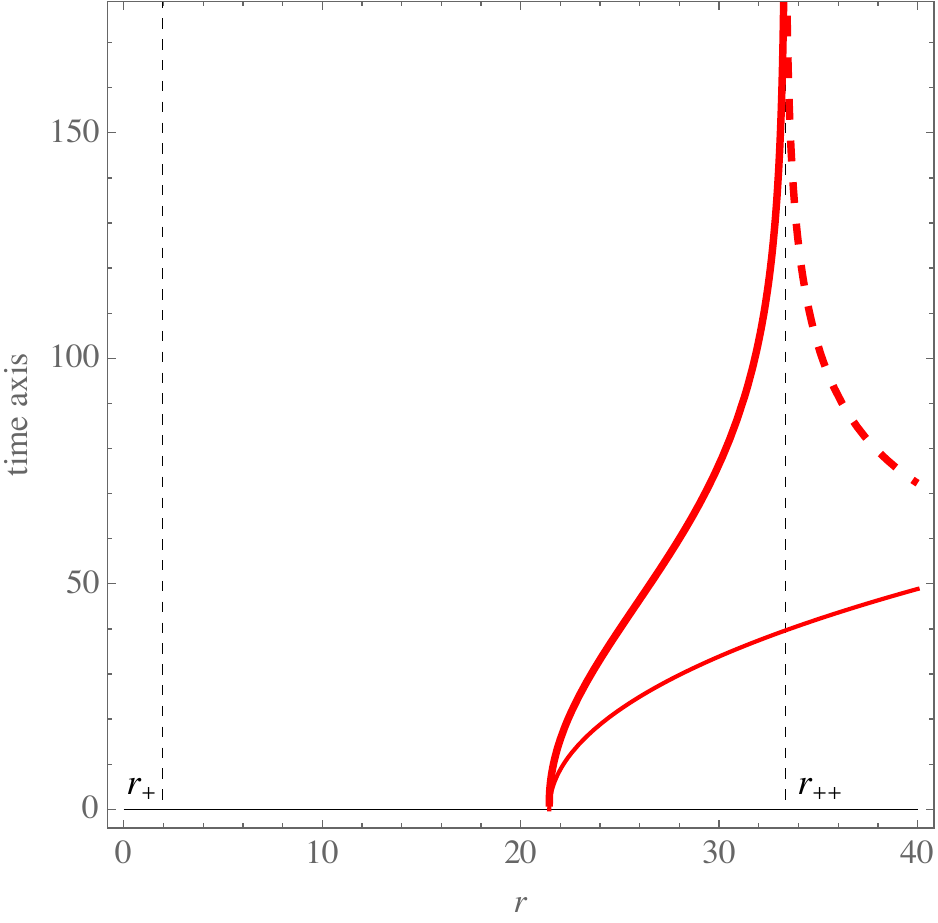}~(a)\qquad\qquad
     \includegraphics[width=7cm]{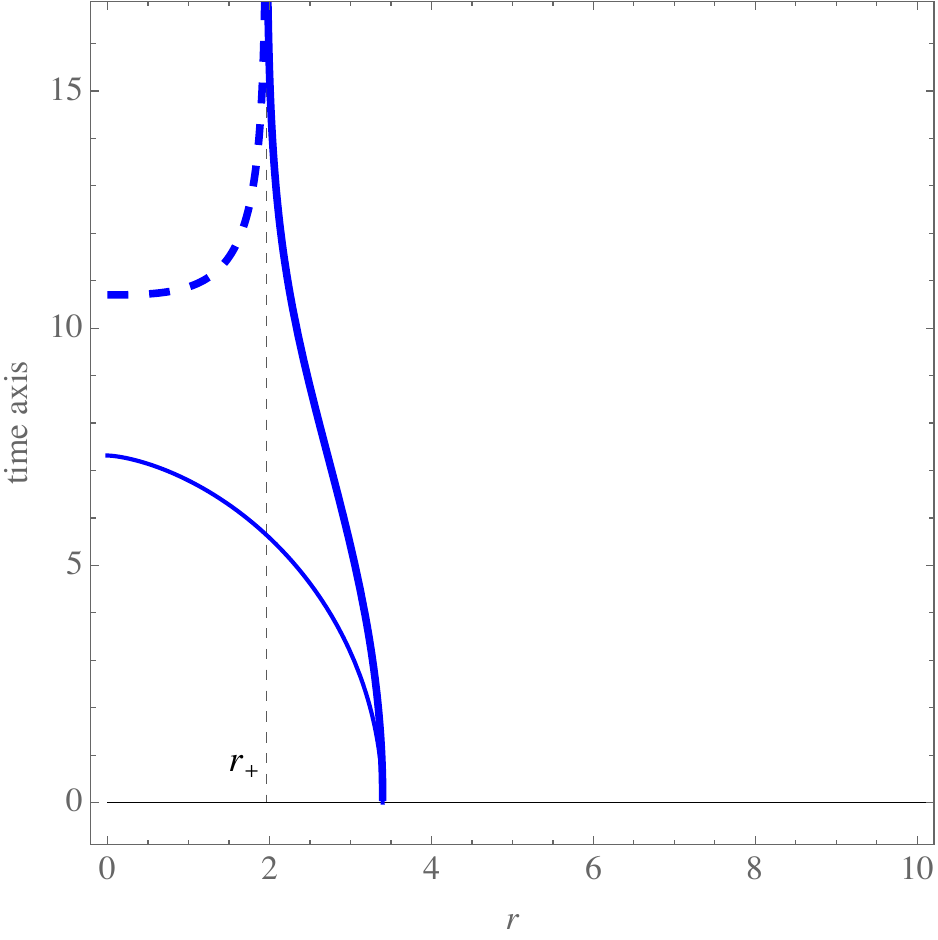}~(b)
    \caption{The plots of (a) the FSFK, and (b) the FSSK, plotted for $E^2=0.4$, $\epsilon=0.02$ and $\tL=3\times10^{-4}$. The corresponding initial points are $d_s$ and $d_f$, as indicated in Fig.~\ref{fig:Vr(r)}. The thin curves show the radial profile of $\tau(r)$, whereas the thick ones correspond to that for $t(r)$.}
    \label{fig:FSFK,FSSK}
\end{figure}
As we can see, the $\tau$-profile crosses the horizons in each of the cases, while this never happens for the $t$-profile, and it shows an asymptotic behavior on the horizons. This highlights the fact that to distant observers, it takes infinite time for infalling particles to pass the horizons. 

\subsection{CROFK and CROSK}

\begin{figure}
    \centering
    \includegraphics[width=8cm]{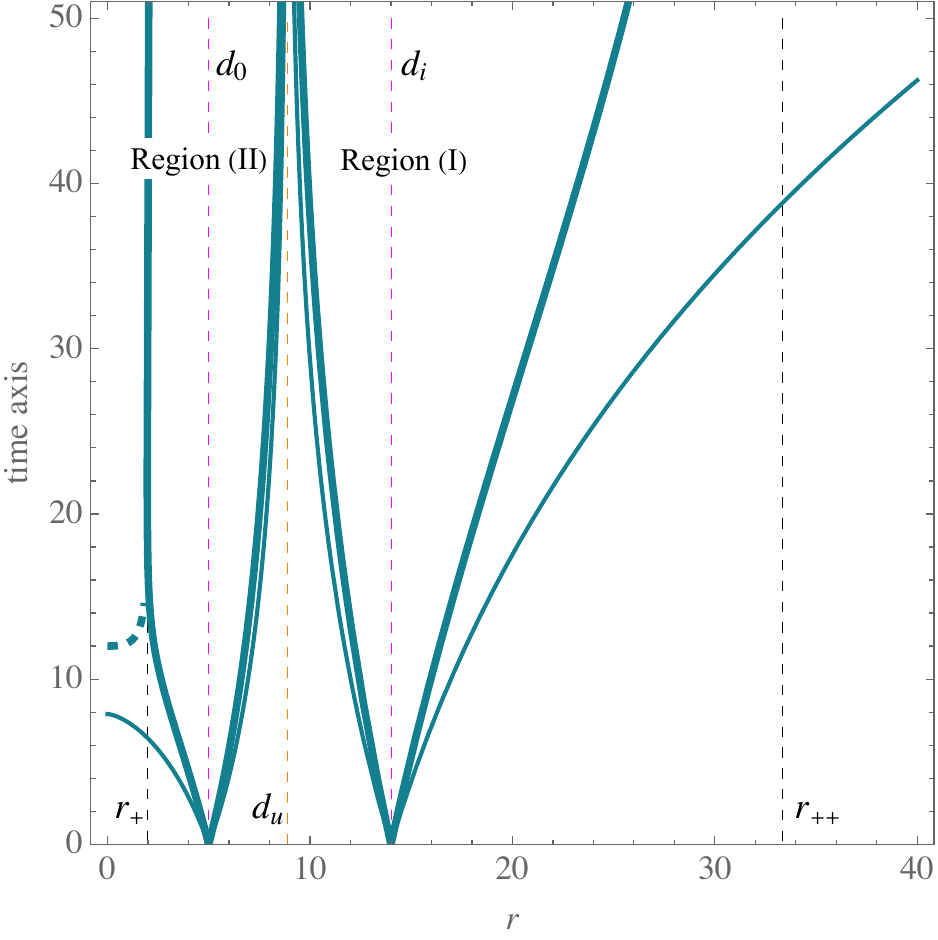}
    \caption{The radial profiles of the temporal
    coordinates for the CROFK and CROSK, plotted for $\epsilon=0.02$ and $\tL=3\times10^{-4}$, which according to Fig.~\ref{fig:Vr(r)} provides $d_u=8.86$ and $E_u^2=0.63$. These orbits can be distinguished in the regions (I) and (II), for which, the initial points of approach are $d_i=15$ and $d_0 = 5$. The thin and thick curves correspond, respectively, to the behaviors of $\tau(r)$ and $t(r)$.}
    \label{fig:CROFSK}
\end{figure}

In this case, the characteristic polynomial in Eq. \eqref{eq:frakp(r)} can be recast as
\begin{equation}
\mathfrak{p}_3(r)=\tL\left(r-d_u\right)^2\left(r-d_1\right),
    \label{eq:frakp(r)-2}
\end{equation}
given $d_u$ in Eq.~\eqref{eq:du}. We can, hence, divide the space into the two regions (I) and (II) which distinguish the fates that occur to the test particles approaching the critical radius $d_u$ from either $d_i$ or $d_0$. Respectively, they correspond to the CROFK and CROSK. Now solving the radial equation of motion \eqref{eq:dotr_rad} for the proper time, these two regions are distinguished by the solutions
\begin{eqnarray}
\tau_\mathrm{I}(r)&=&\pm\frac{1}{\sqrt{\tL}}\left[\tau_A(r,d_u)-\tau_A(d_i,d_u)\right],\label{eq:tau_I}\\
\tau_\mathrm{II}(r)&=&\mp\frac{1}{\sqrt{\tL}}\left[\tau_A(r,d_u)-\tau_A(d_0,d_u)\right],\label{eq:tau_II}
\end{eqnarray}
in which
\begin{equation}
\tau_A(r,d_u) = 2\arctanh\left(\sqrt{\frac{r}{r-d_1}}\,\right)
-2\sqrt{\frac{d_u}{d_u-d_1}}\arctanh\left(\frac{r-d_u+\sqrt{r(r-d_1)}}{\sqrt{d_u(d_u-d_1)}}\right).
    \label{eq:tau_A}
\end{equation}
This is while for the distant observers, the equation of motion \eqref{eq:drdt_rad} provides the solutions
\begin{eqnarray}
t_{\mathrm{I}}(r)&=&\pm\frac{E_u}{\sqrt{\tL^3}}\sum_{n=1}^{4}\varpi_n\left[t_n(r)-t_n(d_i)\right],\label{eq:t_I}\\
t_{\mathrm{II}}(r)&=&\mp\frac{E_u}{\sqrt{\tL^3}}\sum_{n=1}^{4}\varpi_n\left[t_n(r)-t_n(d_0)\right],\label{eq:t_I}
\end{eqnarray}
for the aforementioned regions, where 
\begin{subequations}\label{eq:t_ns}
\begin{align}
    & t_1(r)=\arctanh\left(\sqrt{\frac{r_{++}-d_1}{r_{++}}}\sqrt{\frac{r}{r-d_1}}\right),\\
    & t_2(r)=\arctanh\left(\sqrt{\frac{r_{+}-d_1}{r_{+}}}\sqrt{\frac{r}{r-d_1}}\right),\\
     & t_3(r)=\arctanh\left(\sqrt{\frac{d_u-d_1}{d_u}}\sqrt{\frac{r}{r-d_1}}\right),\\
     & t_4(r)=\arctanh\left(\sqrt{\frac{r_{3}-d_1}{r_{3}}}\sqrt{\frac{r}{r-d_1}}\right),
\end{align}
\end{subequations}
and
\begin{subequations}\label{eq:varpis}
\begin{align}
    & \varpi_1=\frac{2\sqrt{r_{++}^3}}{(r_{++}-r_3)(r_{++}-r_+)(r_{++}-d_u)\sqrt{r_{++}-d_1}},\\
     & \varpi_2=\frac{2\sqrt{r_{+}^3}}{(r_{++}-r_+)(r_{+}-r_3)(d_u-r_+)\sqrt{r_{+}-d_1}},\\
     & \varpi_3=\frac{2\sqrt{d_u^3}}{(r_{++}-d_u)(d_u-r_+)(d_u-r_3)\sqrt{d_u-d_1}},\\
     & \varpi_4=-\frac{2\sqrt{r_3^3}}{(r_{++}-r_3)(r_+-r_3)(d_u-r_3)\sqrt{d_1-r_3}}.
\end{align}
\end{subequations}
The radial profiles of the time coordinates have been plotted in Fig.~\ref{fig:CROFSK}, in the contexts of the CROFK and CROSK, within the discussed regions and based on the initial points of approach.

In this section, we studied the motion of particles with zero initial angular momentum. We classified the orbits and obtained the fully analytical solutions to the equations of motion. On the other hand, the more general types of orbits occur when the particles approach the black hole with non-zero initial angular momentum. Hence, in the next section, we proceed with our discussion by studying angular geodesics.

\section{Angular motion}\label{sec:angular}

\begin{figure}
    \centering
    \includegraphics[width=8.2cm]{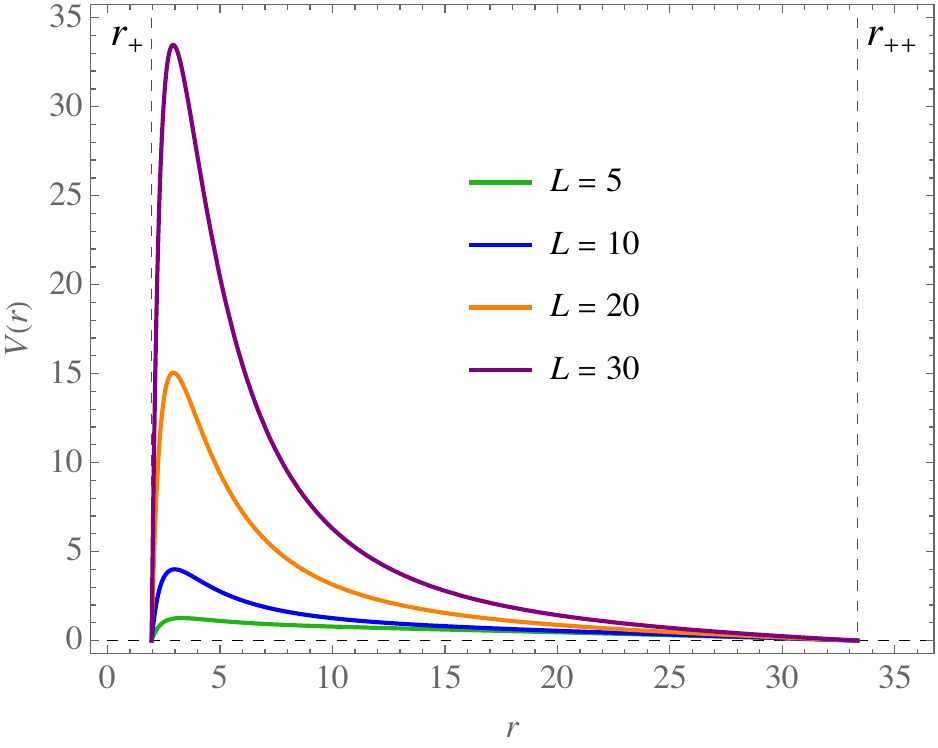}~(a)\qquad
    \includegraphics[width=8.2cm]{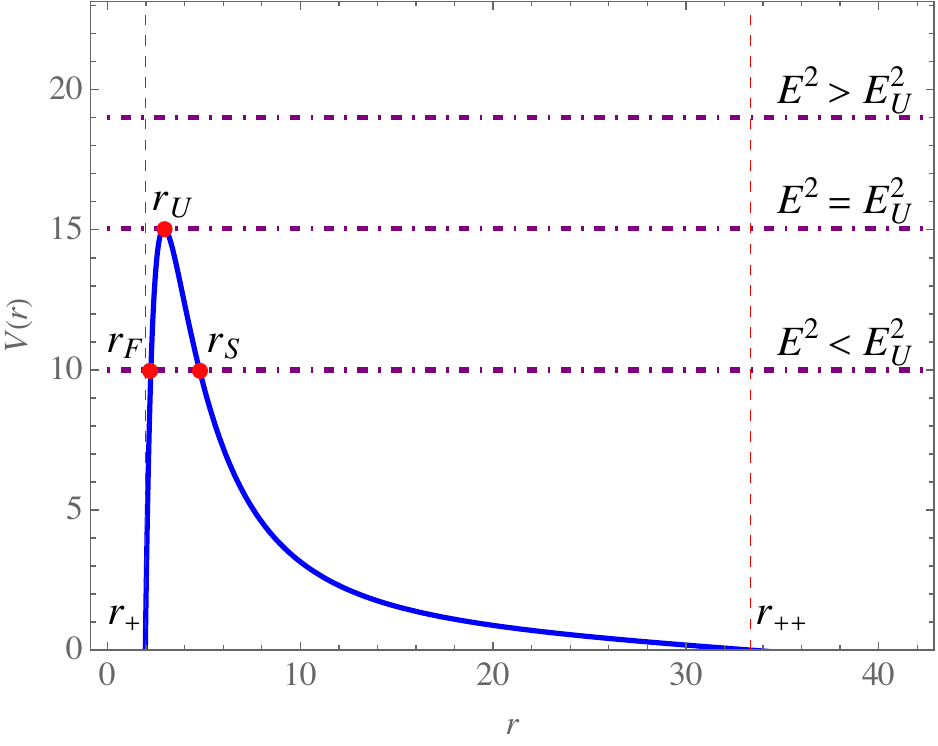}~(b)
    \caption{The radial profiles of the angular effective potential, plotted for $\epsilon=0.02$ and $\tL=3\times 10^{-4}$, and for (a) four different initial angular momenta, and (b) the particular case of $L=20$. Accordingly, the critical radius is $r_U = 2.93$, corresponding to the potential's extremum $V(r_U)=15.05=E_U^2$, and the two turning points $r_S = 4.77$ and $r_F = 2.25$ are encountered for $E^2=10$. }
    \label{fig:V(r)}
\end{figure}

Test particles that approach the black hole with non-zero initial angular momentum (i.e. $L\neq0$), will travel on angular geodesics which are of more diversity and importance. In this section, we perform an analytical study on the different types of angular motion around the SDdS black hole, by solving the equation of motion \eqref{eq:drdphi}. These orbits are classified by means of the effective potential \eqref{eq:V(r)}, whose radial profile has been plotted in Fig. \ref{fig:V(r)}, for different values of the test particle's angular momentum. Each of the profiles possesses a maximum point, at which, the orbits may become unstable. By raising the initial angular momentum, the height of this maximum is increased, and the profile becomes steeper after this point. As indicated in the right panel of Fig. \ref{fig:V(r)}, the orbits may encounter different turning points $r_t$ regarding their initial energy values, that satisfy $E^2=V(r_t)$. According to Fig.~\ref{fig:V(r)}(b), 
circular orbits happen at the maximum, where $r_t=r_U$, and orbits of the first and second kinds (OFK and OSK) occur, respectively, at $r_t=r_S$ and $r_t=r_F$, for which $E^2<E_U^2$. Once $E^2>E_U^2$, the trajectories are captured by the black hole.  
Note that, since the effective potential does not have any minimums,
the SDdS black hole is not capable of forming an accretion disk, which requires the availability of innermost stable circular orbits (ISCO). However, the spirally infalling particles can be detected by means of their direct emission before being devoured into the event horizon.

\subsection{Circular orbits}\label{subsec:CO}

\begin{figure}[h]
    \centering
    \includegraphics[width=8.2cm]{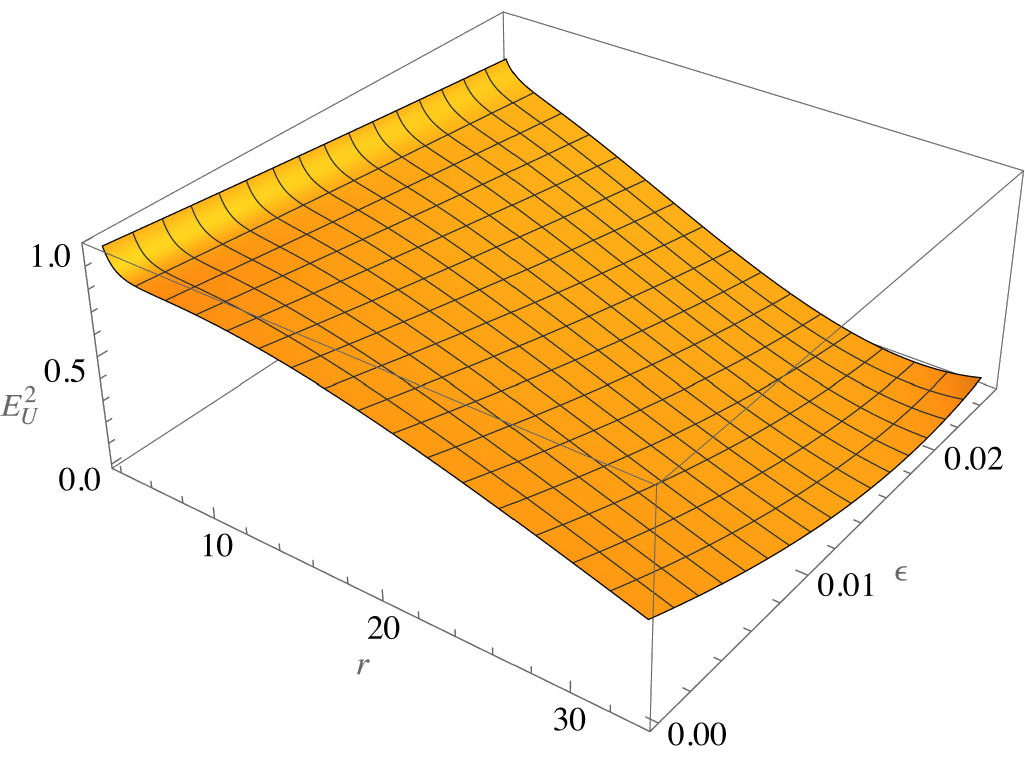}~(a)\qquad
    \includegraphics[width=8.2cm]{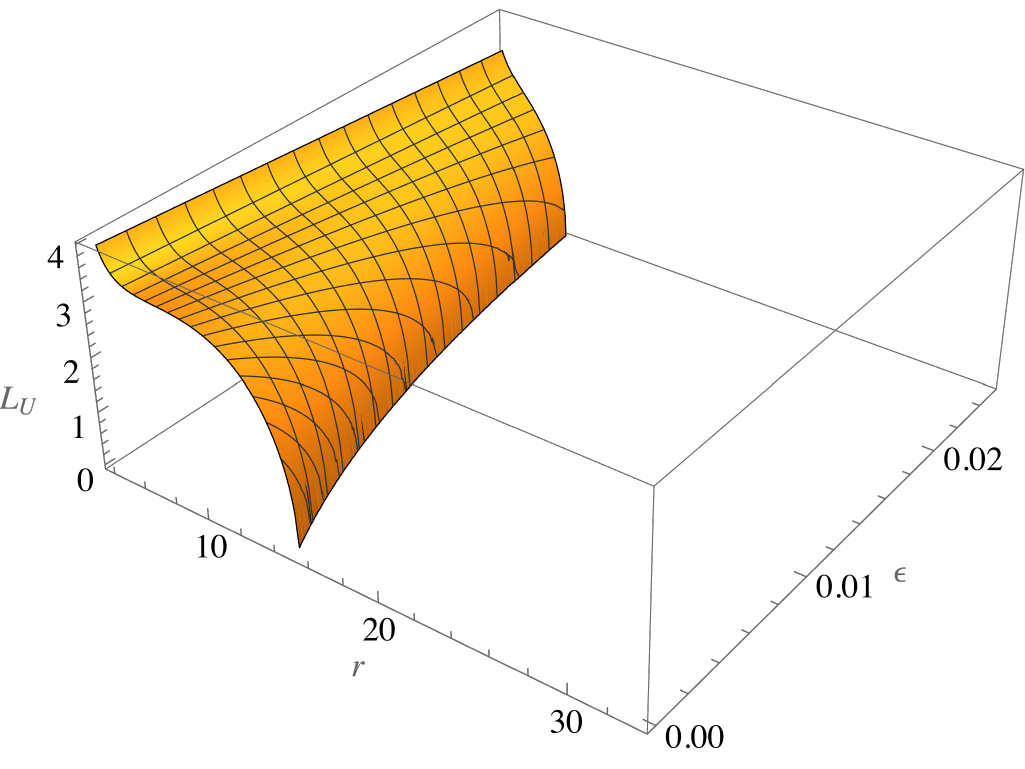}~(b)
    \caption{The behavior of the radial profiles of $E_U^2$ and $L_U$, respecting the variations in the $\epsilon$-parameter plotted for $\tL=3\times 10^{-4}$.}
    \label{fig:EULU3D}
\end{figure}

The  circular orbit occurs when the effective potential reaches its maximum, at which $V'(r)=0$. This equation together with the condition $V(r)=E_U^2$ (or $\dot r=0$ in Eq. \eqref{eq:dotr}), yields
\begin{eqnarray}
&& E_U(r)^2 =\frac{2}{r} \frac{\left[\epsilon r^2-(3 \epsilon+1) r+\tL  r^3+2\right]^2}{ 2 (3 \epsilon+1)r-\epsilon r^2 -6},\label{eq:EU}\\
&& L_U(r) = \sqrt\frac{r^2\left(2\tL r^3-\epsilon r^2-2\right)}{\epsilon r^2-6\epsilon r-2 r+6}.\label{eq:LU}
\end{eqnarray}
In Fig. \ref{fig:EU,LU}, the radial profiles of the above quantities have been shown for the specific case of the effective potential in Fig. \ref{fig:V(r)}(b). As expected, by approaching the critical radius $r_U$, the profiles increase sharply until they reach the values $E_U^2=V(r_U)$ and $L_U=L$ (i.e. the critical energy and the initial angular momentum). In contrast, by receding from $r_U$, the energy decreases and approaches its values at the vicinity of the cosmological horizon, whereas the angular momentum falls rapidly and vanishes at the radial distance of unstable radial orbits, $d_u$.  
\begin{figure}
    \centering
    \includegraphics[width=8.2cm]{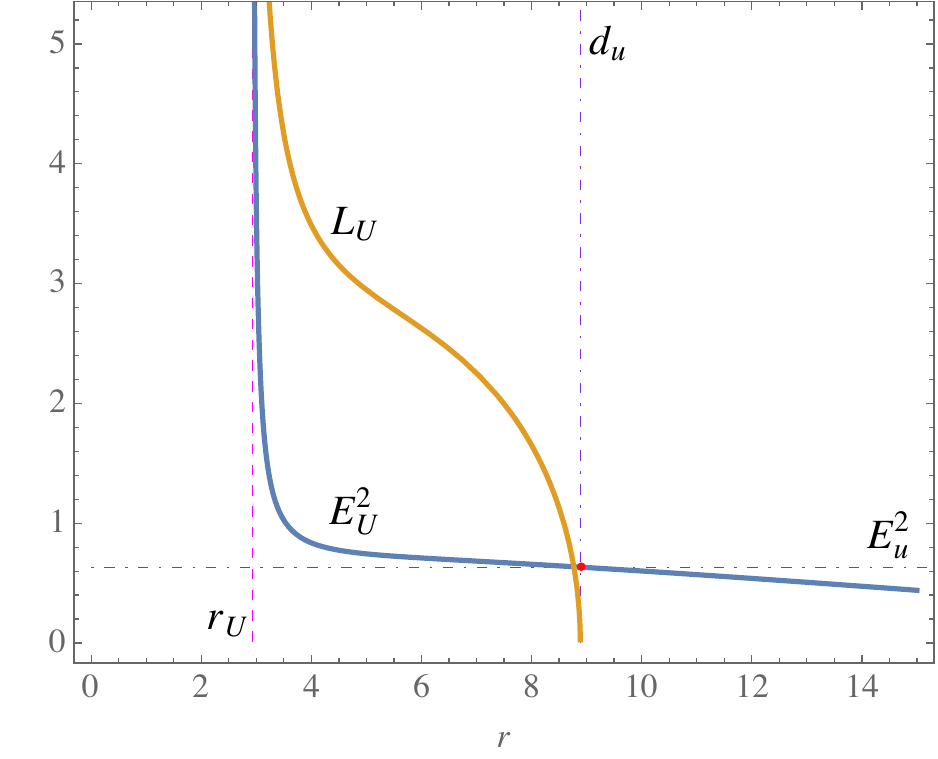}
    \caption{The radial behavior of $E_U^2$ and $L_U$, plotted for $\epsilon=0.02$ and $\tL=3\times 10^{-4}$. By approaching $r_U$, the profiles increase to $E_U^2=V(r_U)=15.05$ and $L_U=20$, which are the essential values of energy and angular momentum that construct the effective potential and its maximum as given in Fig. \ref{fig:V(r)}(b). Moving away from $r_U$, the energy profile falls intensely and then continues to decrease smoothly after passing its value $E_u^2=0.63$ at the radius of critical radial orbits $d_u=8.86$, where the angular momentum vanishes.}
    \label{fig:EU,LU}
\end{figure}
Furthermore, to show the dependence of the above profiles on the variations in the running parameter $\epsilon$, in Fig. \ref{fig:EULU3D}, we have done three-dimensional plots of $E_U^2(r)$ and $L_U(r)$ for the same range of $\epsilon$ as in Fig. \ref{fig:f(r)}.
%

\subsubsection{Stability of the orbits}\label{subsubsec:stability}

\begin{figure}
    \centering
    \includegraphics[width=8.2cm]{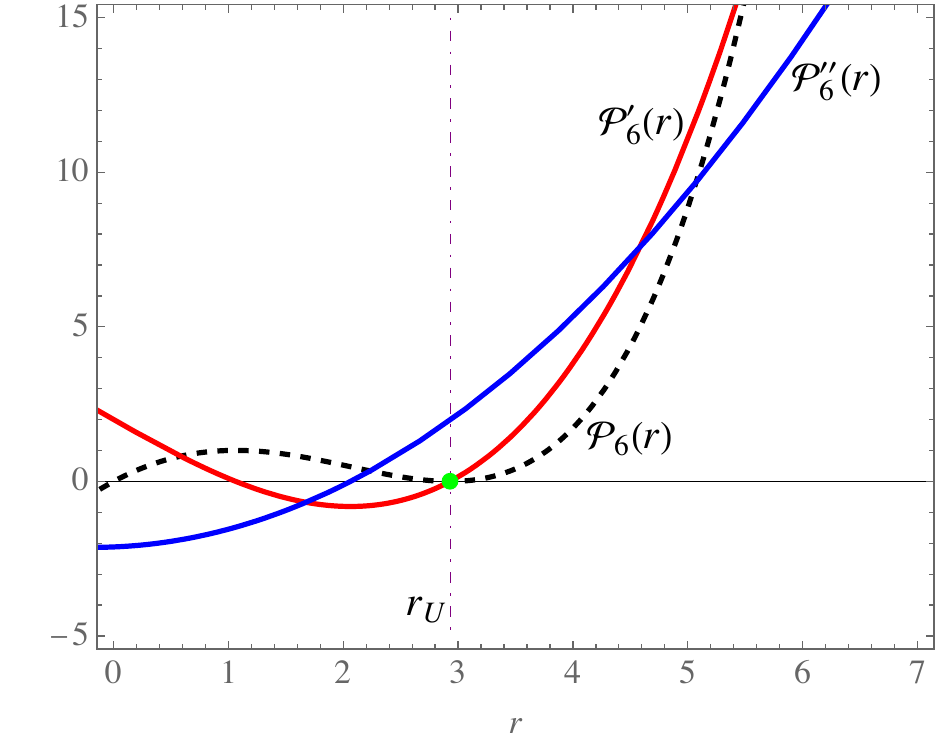}
    \caption{The behavior of $\mathcal{P}_6(r)$ and its first two derivatives plotted for $L=20$, $\epsilon=0.02$ and $\tL=3\times 10^{-4}$. At the turning point $r_U$, we have $\mathcal{P}_6(r)=0=\mathcal{P}_6'(r)$ (the green dot), which is the condition for circular orbits. These orbits are, however, stable since $\mathcal{P}_6''(r_U)>0$.}
    \label{fig:P6}
\end{figure}

\begin{figure}
    \centering
    \includegraphics[width=8.2cm]{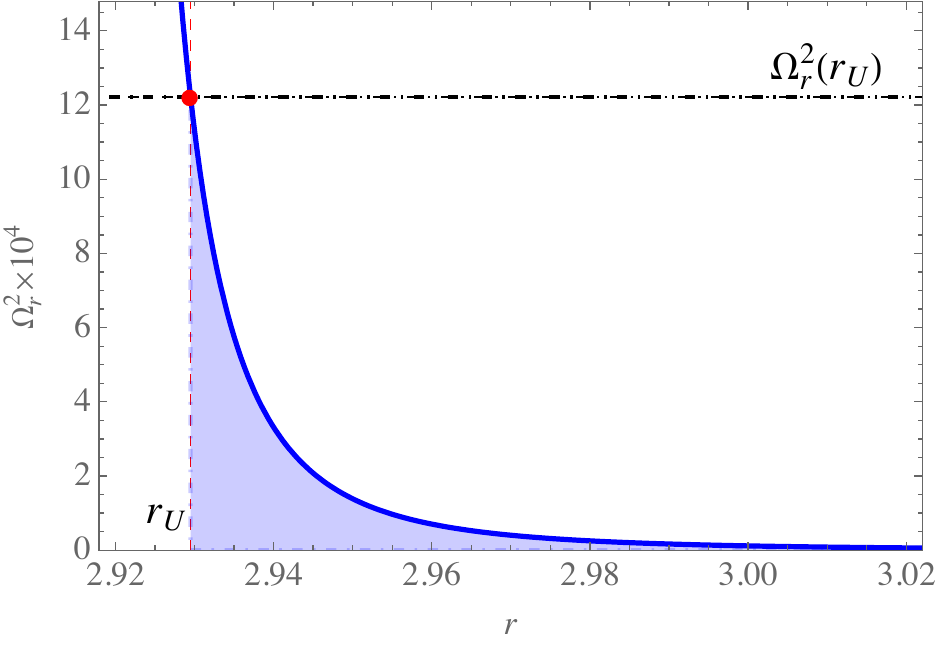}
    \caption{The radial profile of $\Omega_r^2$ plotted for $\epsilon=0.02$ and $\tL=3\times 10^{-4}$. The red dot shows the value of frequency at the radius of circular orbits $r_U$. The non-zero values of $\Omega_r^2$ lie within the domain $r_U\leq r<3$ (the shaded region).}
    \label{fig:Omegar2}
\end{figure}

Let us rewrite the equation of motion \eqref{eq:drdphi} as 
\begin{equation}
\left(\frac{\ed r}{\ed\phi}\right)^2=\frac{\mathcal{P}_6(r)}{L^2},
    \label{eq:drdphi-new}
\end{equation}
in which, the characteristic polynomial is given as
\begin{equation}
\mathcal{P}_6(r) = \tL r^6+\epsilon r^5+\left(L^2\tL+E^2-3\epsilon-1\right)r^4+\left(2-\epsilon L^2\right)r^3-(1+3\epsilon) L^2 r^2+2L^2 r.
\label{eq:P6(r)}    
\end{equation}
This way, the turning points of the angular motion are determined by means of the equation $\mathcal{P}_6(r)=0$. Accordingly, the orbits can become circular at a turning point, where $\mathcal{P}_6(r)=\mathcal{P}_6'(r)=0$, and be marginally stable at that point once the extra condition $\mathcal{P}_6''(r)=0$ is satisfied. Hence, the circular orbits are stable (unstable) when $\mathcal{P}_6''(r)>0$ ($\mathcal{P}_6''(r)<0$). In Fig. \ref{fig:P6}, the behavior of the characteristic polynomial and its differentials has been plotted at the vicinity of the radius of circular orbits $r_U$, for the specific case of Fig. \ref{fig:V(r)}(b). As it can be discerned from the diagram, at the vicinity of the radius of circular obits we have $\mathcal{P}_6''(r)>0$, which indicates that the circular orbits at this radius have some amount of stability. 
This stability stems from the curve width at the tip of the effective potential, which is non-zero for all values of the test particle's initial angular momentum. However, to find the extent, to which the circular orbits are stable, one needs to calculate the sensitivity of the circular orbits to perturbations along the radial axis. This way, a limit can be identified, beyond which, the circular orbits become unstable. Such limit can be obtained in the context of \textit{epicyclic frequency} $\Omega_r$, which is the frequency of oscillations of circularly orbiting particles along the radial direction \cite{Abramowicz_epicyclic:2005} (see also the review in Ref. \cite{Abramowicz_foundations:2013}). In the case of the SDdS black hole, this frequency can be expressed as \cite{Rayimbaev_dynamics;2021}
\begin{equation}
\Omega_r^2(r)=-\frac{1}{2g_{rr}}{V}''(r),
\label{eq:epicvr}    
\end{equation}
which by means of Eqs. \eqref{eq:LU} and \eqref{eq:V(r)}, yields
\begin{multline}
\Omega_r^2(r)=\frac{8 r \left\{-3 \tL ^2 \left[r (2 r-15)+30\right] r^5+2 \tL  \left[r \left(
r^2-9r+27\right)-18\right] r^2+(r-6)^2\right\}-288}{\left(6-2 r\right)^3 r^2 \left(\tL  r^3-r+2\right)}\\
+\frac{\epsilon}{{2 (r-3)^4 r \left(\tL  r^3-r+2\right)^2}}
\left[10 \tL ^3 r^{12}-120 \tL ^3 r^{11}+513 \tL ^3 r^{10}-16 \tL ^2 r^{10}-810 \tL ^3 r^9+234 \tL ^2 r^9\right.\\
\left.-1305 \tL ^2 r^8-2 \tL  r^8+3294 \tL ^2 r^7+20 \tL  r^7-2916 \tL ^2 r^6-57 \tL  r^6-54 \tL  r^5\right.\\
\left.+2 r^5+216 \tL  r^4-31 r^4+218 r^3-708 r^2+1080 r-648\right]+\mathcal{O}\left(\epsilon^2\right).
    \label{eq:Omegar2}
\end{multline}
The radial behavior of the radial epicyclic frequency for the circular orbits at the radius $r_U$ in the effective potential in Fig.~\ref{fig:V(r)}(b), has been demonstrated in Fig.~\ref{fig:Omegar2}.
As it can be observed from the figure, the frequency falls rapidly from its high values at $r_U$, and tend to zero as we recede from it. As long as $\Omega_r^2\neq0$, we can expect stability of circular orbits at the vicinity of $r_U$. In this sense, the stability domain for this particular case is within $r_U\leq r<3$. Passing this region, the particles do not travel on stable orbits and escape from the black hole (see the forthcoming sections).

\subsection{OFK and the scattering zone}\label{subsec:OFK}

Once the test particles are subjected to the condition $E^2<E_U^2$, they encounter the two turning points $r_S$ and $r_F$, while approaching the black hole (see Fig. \ref{fig:V(r)}b). In this sense, the characteristic polynomial \eqref{eq:P6(r)} can be recast as
\begin{equation}
\mathcal{P}_6(r) = \tL r\left(r-r_S\right)\left(r-r_F\right)\left(r-r_4\right)\left(r-r_5\right)\left(
r-r_5^*\right),
    \label{eq:P6(r)-def}
\end{equation}
in which $r_4<0$ and $r_5\in\Bbb{C}$. The test particles approaching from $r_S$, experience a hyperbolic motion and then escape from the black hole. This scattering phenomenon occurs in the context of the OFK. However, to obtain the explicit solution to the angular equation of motion \eqref{eq:drdphi-new}, we encounter the inversion of the hyper-elliptic integral
\begin{equation}
    \phi-\phi_0=L\int_{r}^{r_S}\frac{\ed r}{\sqrt{\mathcal{P}_6(r)}},
    \label{eq:phi(r)-int}
\end{equation}
with $\phi_0$ being the initial azimuth angle, which confronts us with a special case of the Jacobi inversion problem. However, before proceeding with the calculation of the inversion, let us provide an analytical expression for $\phi(r)$, by doing a direct integration of Eq. \eqref{eq:phi(r)-int}. This solution is of importance once the deflection angle of scattered particles is concerned. Now considering the expression \eqref{eq:P6(r)-def} and using the method given in appendix \ref{app:A}, we obtain the analytical solution
\begin{equation}
\phi(r)=\phi_0-\frac{L r_S}{2\sqrt{\tL\mathscr{T}^4}}\left(1-\frac{r_S}{r}\right)^2 F_D^{(5)}
\left(
2,\frac{1}{2},\frac{1}{2},\frac{1}{2},\frac{1}{2},\frac{1}{2};3;\delta_1,\delta_2,\delta_3,\delta_4,1-\frac{r_S}{r}
\right),
    \label{eq:phi(r)-OFK}
\end{equation}
where $\mathscr{T}^4=r_S r_F^{-2}r_4^{-1} |r_5|^2(r_S-r_F)(r_F-r_4)(r_S-r_5)(r_S-r_5^*)$, and 
\begin{subequations}
\begin{align}
    & \delta_1 = \frac{r_S}{r_S-r_F},\\
    & \delta_2 = \frac{r_S}{r_S-r_4},\\
    & \delta_3 = \frac{r_S}{r_S-r_5} = \delta_4^*.
\end{align}
\label{eq:deltai}
\end{subequations}
Note that, since the integral equation \eqref{eq:phi(r)-int} is generically hyper-elliptic, it cannot be solved explicitly in terms of common elliptic integrals. Nevertheless, under some circumstances, it could be reduced to a degenerate hyper-elliptic integral, and be solved in the same way as for the equation of motion \eqref{eq:dotr_rad} (see appendix \ref{app:B}).

\subsubsection{The scattering angle}\label{subsubsec:scatteringAngle}

\begin{figure}[t]
    \centering
    \includegraphics[width=8.2cm]{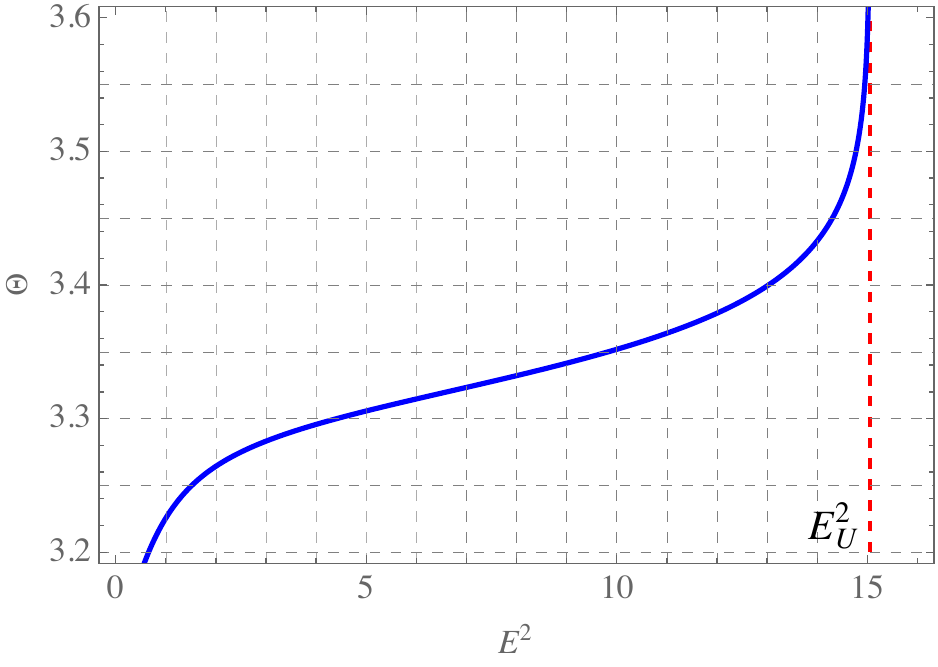}
    \caption{The change of $\Theta$ versus the variations in $E^2$, plotted for $L=20$, $\epsilon=0.02$ and $\tL=3\times 10^{-4}$, in the domain $0\leq E^2\leq E_U^2=15.05$ and for $r_\mathrm{O}=28$. As expected, the scattering angle has a fixed value at the vicinity of the cosmological horizon, where the energy vanishes, and then diverges by approaching the energy of circular orbits.}
    \label{fig:ThetaE2}
\end{figure}

After reaching the point of closest approach $r_S$, particles on the OFK are scattered by the black hole. The same as the case of deflection of light in black hole spacetimes, for an observer located at the radial distance $r_\mathrm{O}$, the deflection (scattering) of massive particles is done under the scattering angle $\Theta = 2\phi_\mathrm{O}-\pi$ \cite{Tsukamoto:2017}, in which $\phi_\mathrm{O}=\phi(r_\mathrm{O})$ is obtained by means of Eq. \eqref{eq:phi(r)-OFK}. In Fig.~\ref{fig:ThetaE2}, we have shown the change of the scattering angle $\Theta$ with respect to the variations in the energy in the domain $0<E<E_U^2$. In order to generate this plot, a set of $(E_S^2,r_S)$ pairs was generated in the context of the effective potential in Fig.~\ref{fig:V(r)}(b), and then by exploiting Eq.~\eqref{eq:phi(r)-OFK}, the values of $\phi_\mathrm{O}$ and their corresponding scattering angles were calculated. 
%

\subsection{Analytical solutions for the orbits}\label{subsec;analyticalSolutions}

\begin{figure}
    \centering
    \includegraphics[width=9cm]{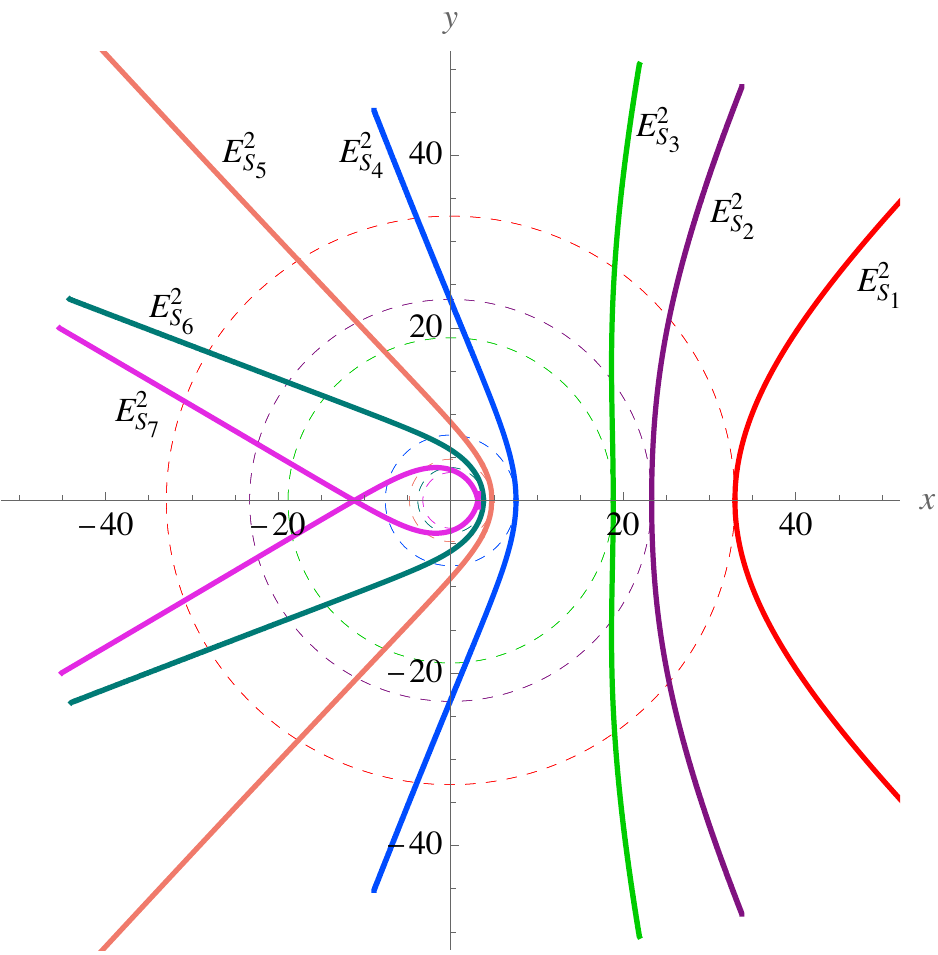}
    \caption{The OFK plotted for $\epsilon=0.02$, $\tL=3\times 10^{-4}$, $\mathscr{L}=0.0025$, $\phi_0=0$, and different energy levels corresponding to particle scattering, in accordance with the effective potential in Fig.~\ref{fig:V(r)}(b). These chosen values are
    $E_{S_1}^2=0.02, E_{S_2}^2=0.6, E_{S_3}^2=1, E_{S_4}^2=5, E_{S_5}^2=10, E_{S_6}^2=13$ and $E_{S_7}^2=14.7$, for which the scattering happens at the radii $r_{S_1}=32.95, r_{S_2}=23.30, r_{S_3}=18.85, r_{S_4}=7.57, r_{S_5}=4.77, r_{S_6}=3.80$ and  $r_{S_7}=3.22$ (the dashed circles with the same colors as the trajectory curves).}
    \label{fig:OFK}
\end{figure}

\begin{figure}
    \centering
    \includegraphics[width=9cm]{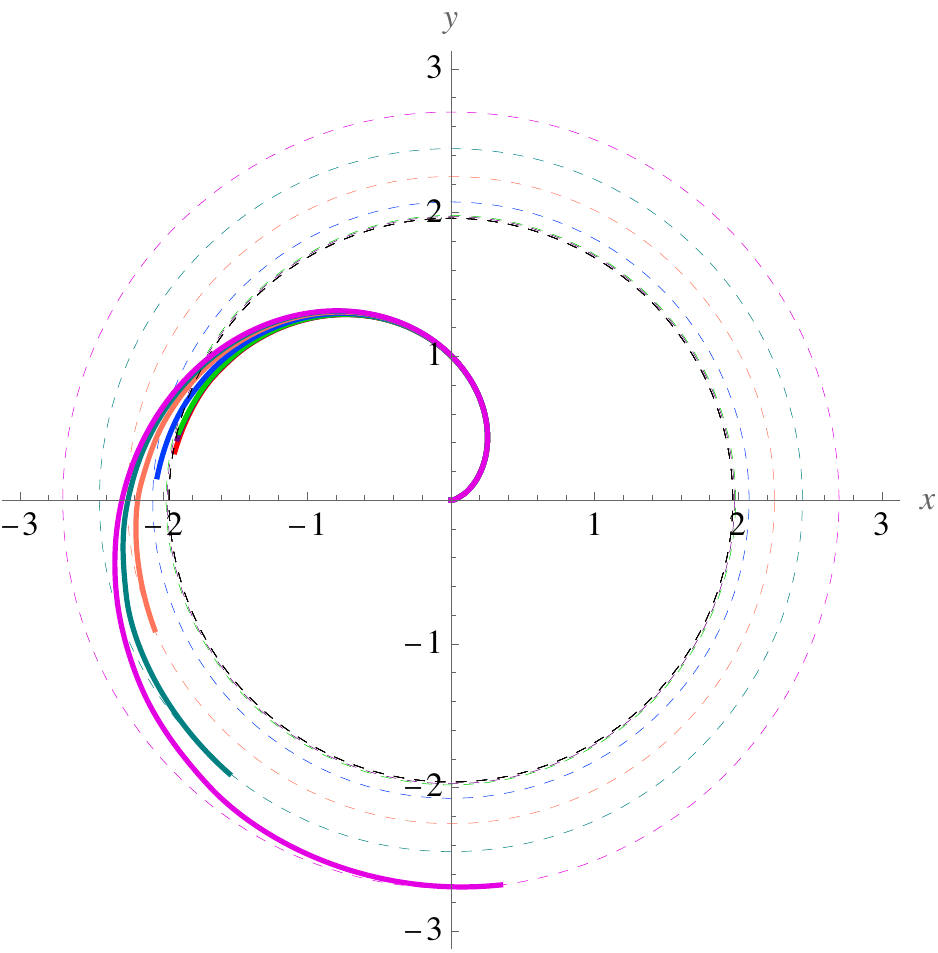}
    \caption{The OSK plotted for $\epsilon=0.02$, $\tL=3\times 10^{-4}$, $\mathscr{L}=0.0025$, $\phi_0=0$, and the same color-coding of the different energy levels as in Fig.~\ref{fig:OFK}. The initial radii of approach have been indicated by dashed circles of the same colors of the orbits, which are $r_{F_1}=1.96, r_{F_2}=1.97, r_{F_3}=1.98, r_{F_4}=2.10, r_{F_5}=2.25, r_{F_6}=2.45$ and  $r_{F_7}=2.70$. The black dashed circle corresponds to $r_+$.}
    \label{fig:OSK}
\end{figure}

\begin{figure}
    \centering
    \includegraphics[width=9cm]{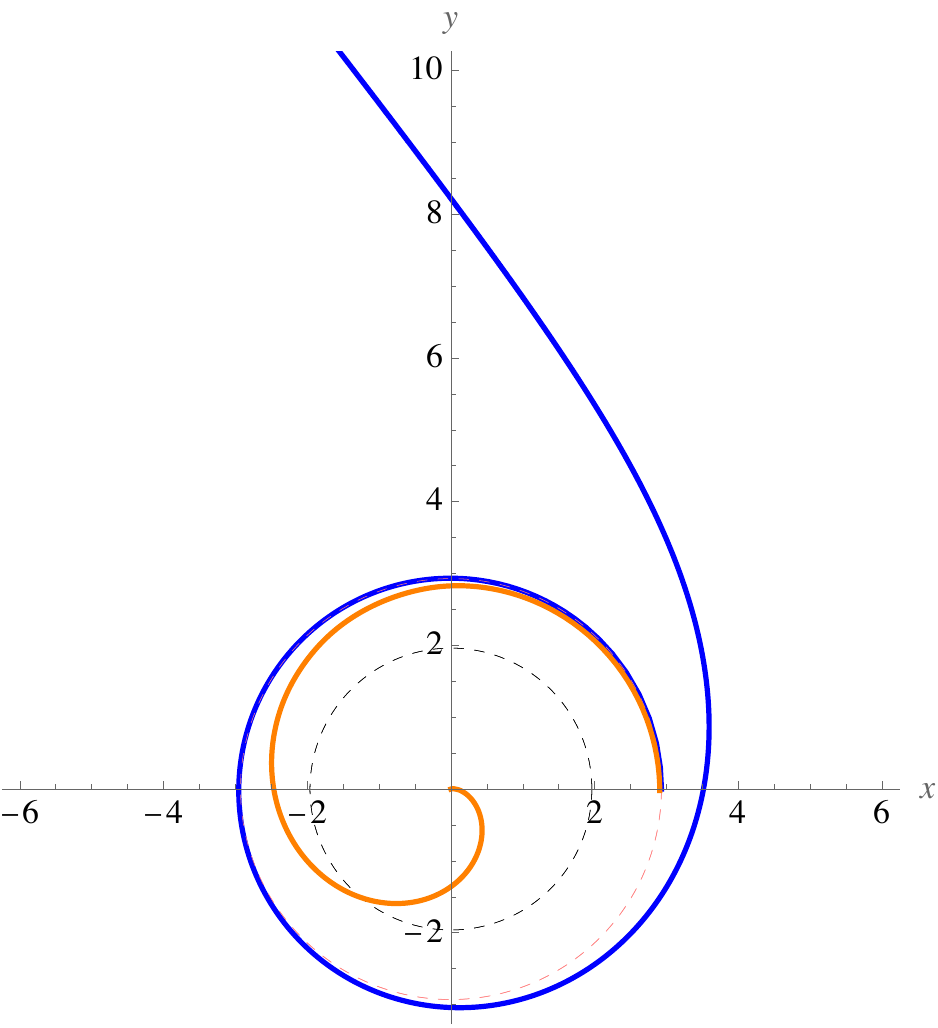}
    \caption{The COFK (blue color) and COSK (orange color) plotted for $\epsilon=0.02$, $\tL=3\times 10^{-4}$, $\mathscr{L}=0.0025$, and $\phi_0=0$. The pink and black dashed circles indicate, respectively, the radii $r_U$ and $r_+$.}
    \label{fig:COFSK}
\end{figure}

\begin{figure}
    \centering
    \includegraphics[width=9cm]{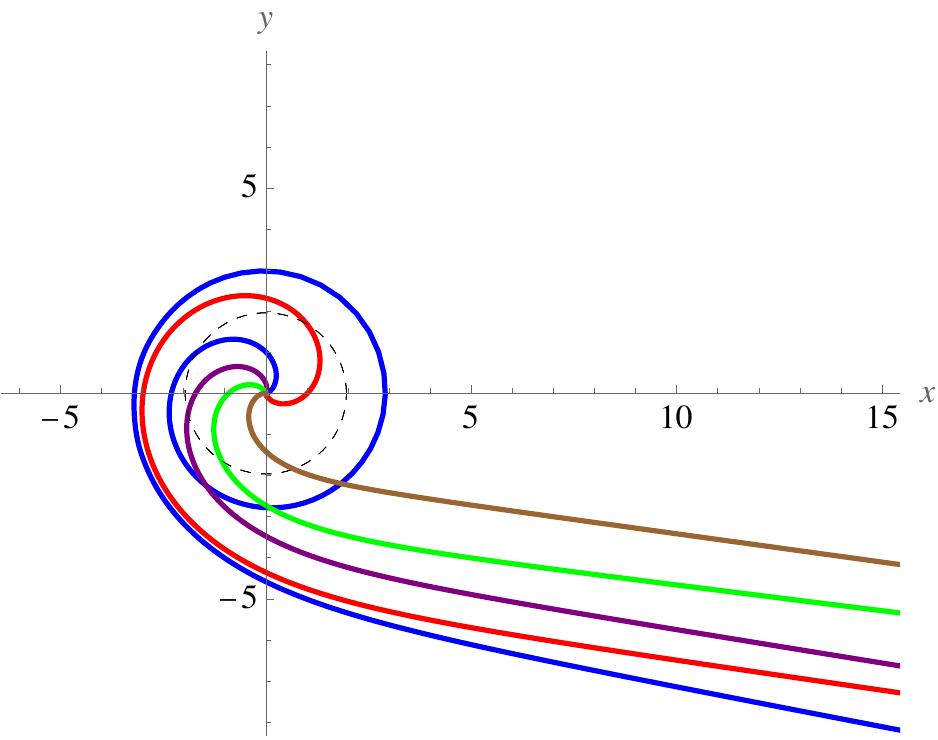}
    \caption{The captured geodesics plotted for $\epsilon=0.02$, $\tL=3\times 10^{-4}$, $\mathscr{L}=0.0025$, $\phi_0=0$, and $r_i=15$. From bottom to top, the curves correspond to $E= E_U+0.001, E_U+0.1, E_U+1, E_U+2,$ and $E_U+6$. The dashed circle in the middle corresponds to $r_+$. }
    \label{fig:capture-sim}
\end{figure}

To present a full study of the possible orbits of particles around the SDdS black hole, here we proceed with constructing a set of exact analytical solutions to the angular equation of motion, which is capable of describing all kinds of orbits in the spacetime geometry. Applying the change of variable $r=1/u$ to Eq. \eqref{eq:drdphi-new}, we get 
\begin{equation}
\phi-\phi_0 = \int_{u_0}^{u}\frac{u\,\ed u}{\sqrt{\mathcal{P}_5(u)}},
    \label{eq:phi(r)-int-1}
\end{equation}
in which $u_0$ corresponds to an initial point of approach located at $(r_0,\phi_0)$, and 
\begin{equation}
\mathcal{P}_5(u)=\tL\mathscr{L}+\epsilon\mathscr{L} u+\left[\tL+\left(E^2-3\epsilon-1\right)\mathscr{L}\right]u^2+
\left(2\mathscr{L}-\epsilon \right)u^3-\left(1+3\epsilon\right)  u^4+2 u^5,
    \label{eq:p5(u)}
\end{equation}
where we have defined $\mathscr{L}=1/L^2$. The Eq. \eqref{eq:phi(r)-int-1} includes a hyper-elliptic integral, for the inverse of which one must use abelian modular functions of genus two. A rigorous method of dealing with such problems was introduced by Riemann to study the singularities of algebraic curves on a homology surface \cite{Riemann:1857}. He also introduced the concept of Riemannian theta functions \cite{Riemann+1866+161+172}, which have been used to solve the raised Jacobi inversion problems. Such functions have also proved very useful in mathematics and theoretical physics. The usefulness of modular forms in general relativity and the applications of genus-2 Riemannian theta functions to the hyper-elliptic integrals arising from the geodesic equations in cosmological-constant-induced spacetimes, were first studied in Refs. \cite{kraniotis_general_2002,kraniotis_compact_2003,kraniotis_precise_2004,kraniotis_frame_2005}, and then in Refs.~\cite{hackmann_complete_2008,hackmann_geodesic_2008,hackmann_analytic_2009,hackmann_complete_2010}, where the Jacobi inversion problem is approached by means of the Riemann surfaces of genus two and higher (see also Ref. \cite{enolski_inversion_2011}). In fact, the square root of the integrand of Eq. \eqref{eq:phi(r)-int-1} has two branches and hence, it is not well defined on the complex plane. Furthermore, we must bear in mind that the inverse solution $u(\phi)$ should not depend on the path of integration \cite{hackmann_geodesic_2008}. In this sense, if
\begin{equation}
\omega=\oint_\gamma\frac{u\,\ed u}{\sqrt{\mathcal{P}_5(u)}},
    \label{eq:omega_1}
\end{equation}
is valid for the integration path $\gamma$, we then expect that
\begin{equation}
\phi-\phi_0-\omega=\int_{u_0}^{u}\frac{u\,\ed u}{\sqrt{\mathcal{P}_5(u)}},
    \label{eq:phi(r)-int-2}
\end{equation}
to be valid as well. Accordingly, the solution must respect the condition $u(\phi)=u(\phi-\omega)$ for all $\omega\neq0$. Now defining the algebraic curve $y^2=\mathcal{P}_5(u)$, homologous to a genus-2 Riemann surface, one can then introduce the holomorphic
\begin{equation}
\ed \zeta_1 = \frac{\ed x}{\sqrt{\mathcal{P}_5(x)}},\qquad
\ed\zeta_2 = \frac{x \,\ed x}{\sqrt{\mathcal{P}_5(x)}},
    \label{eq:dzetai}
\end{equation}
and meromorphic 
\begin{equation}
\ed\rho_1 = \frac{\left(2\mathscr{L}-\epsilon \right)x-2\left(1+3\epsilon\right) x^2+6x^3}{4\sqrt{\mathcal{P}_5(x)}}\ed x,\qquad 
\ed\rho_2=\frac{4x^2}{4\sqrt{\mathcal{P}_5(x)}}\ed x,
\label{eq:drhoi}
\end{equation}
differentials in accordance with the expression in Eq. \eqref{eq:p5(u)}, and based on the definitions given in Ref. \cite{Buchstaber:1997}. We also introduce the real \begin{equation}
2\omega_{ij} = \oint_{a_j}\ed \zeta_i,\qquad 
2\eta_{ij} = \oint_{a_j}\ed \rho_i,
    \label{eq:realhalfPer}
\end{equation}
and imaginary
\begin{equation}
2\bar{\omega}_{ij} = \oint_{b_j}\ed\zeta_i,\qquad 
2\bar{\eta}_{ij} = \oint_{b_j}\ed\rho_i,
    \label{eq:imaghalfPer}
\end{equation}
half-period matrices on the homology basis $(a_i,b_i)$ of the Riemann surface. Together, the above quantities generate the symmetric period matrices of the first and second kinds, given respectively as $(2\bm{\omega},2\bar{\bm{\omega}})$ and $(2\bm{\eta},2\bar{\bm{\eta}})$. Having these information in hand, the analytical solution to the inversion of the integral equation \eqref{eq:phi(r)-int-2} is obtained as \cite{hackmann_geodesic_2008,enolski_inversion_2011}
\begin{equation}
u(\phi) = -\frac{\sigma_1}{\sigma_2}(\phi_\sigma),
    \label{eq:u(phi)-0}
\end{equation}
in which $\sigma_i$ represents the $i$th derivative of the 2-variable Kleinian sigma function
\begin{equation}
\sigma(\bm{z}) = \mathcal{C} e^{\bm{z}^{t}\cdot\bm{\mathcal{K}}\cdot\bm{z}}\,\vartheta\left[\begin{matrix}
         \bm{q}  \\
         \bm{q}' 
    \end{matrix}
    \right]\left(2\bm{\omega}^{-1}\cdot\bm{z};\bm{T}\right),
    \label{eq:sigma_def}
\end{equation}
which is expressed in terms of the genus-2 Riemannian theta function
\begin{equation}
    \vartheta\left[\begin{matrix}
         \bm{q}  \\
         \bm{q}' 
    \end{matrix}
    \right] \left(\bm{z};\bm{T}\right)= \sum_{\bm{m}\in\Bbb{Z}^2} e^{\mathrm{i}\pi\left(\bm{m}+\bm{q}\right)^{t}\cdot\left[\bm{T}\cdot(\bm{m}+\bm{q})+2\bm{z}+2\bm{q}'\right]},
\end{equation}
with characteristics $\bm{q} = (0,1/2)^t$ and $\bm{q}'=(1/2,1/2)^t$,
where $\bm{T}=\bm{\omega}^{-1}\cdot\bar{\bm{\omega}}$ is the symmetric Riemann matrix, $\bm{\mathcal{K}}=\bm{\eta}\cdot(2\bm{\omega})^{-1}$, and the vector of Riemann constants is given as $\bm{K} = \bm{q}+\bm{q}'\cdot\bm{T}$. In the above relations, the sign $\cdot$ indicates the common matrix product. Also, the constant $\mathcal{C}$ has certain properties and can be obtained explicitly \cite{Buchstaber:1997}. Moreover, $\phi_\sigma = \left(\mathscr{F}(\phi-\phi_{\mathrm{in}}),\phi-\phi_{\mathrm{in}}\right)^t$ with $\phi_\mathrm{in} = \int_{u_0}^\infty\frac{u\,\ed u}{\sqrt{\mathcal{P}_5(u)}}$, is a one-dimensional divisor. This sigma divisor can be obtained by means of the extra condition $\sigma(\phi_\sigma) = 0$, which identifies the function $\mathscr{F}$. Finally, the angular profile of the radial coordinate is obtained as
\begin{equation}
r(\phi) = -\frac{\sigma_2}{\sigma_1}(\phi_\sigma).
    \label{eq:r(phi)-0}
\end{equation}
In the above solution, the functions $\sigma_i$ depend on the parameters $\bm{\omega}, \bm{\eta}, \bm{T}$ and  $\phi_\sigma$, as well as the characteristic polynomial $\mathcal{P}_5(u)$. Since this solution is valid in all regions of the spacetime, it can be applied to simulate every kind of particle orbits which are allowed by the effective potential. In Fig.~\ref{fig:OFK}, this solution has been used in the domain $E
^2<E_U^2$, to simulate the OFK for the scattered particles.
As it can be inferred from the diagram, the trajectories are of hyperbolic form in the equatorial plane, and the more the turning points recede from the radius of circular orbits $r_U$, the particles have more tendency to travel on repulsive geodesics. On the other hand, the trajectories become attractive when the turning point approaches $r_U$. For the particular case of $E_{S_7}^2$ in the figure, as expected, the particles approach the circular orbits, but still they escape the black hole. This behavior is closely related to that for the critical orbits. As discussed earlier, the same energy levels produce another turning point $r_F$ on the effective potential, from which, the test particles can only travel on the OSK and be captured by the black hole. In Fig.~\ref{fig:OSK}, the energy level choices of Fig.~\ref{fig:OFK} have been adopted to simulate the OSK on the SDdS black hole.  
These two kinds of orbits, in fact, confine the orbits that occur at the vicinity of the potential's extremum, and are termed as the critical orbits. If the particles with $E^2=E_U^2$ approach this extremum from the radial distances $r_i\gtrsim  r_U$, they finally escape the black hole after performing circular orbits at the radius $r_U$. Such particles travel on the critical orbit of the first kind (COFk). On the other hand, particles of the same energy will travel on the critical orbit of the second kind (COSK), when they approach the extremum form the distances $r_i\lesssim r_U$, and they finally fall onto the event horizon. In Fig.~\ref{fig:COFSK}, these two orbits have been shown together to compare their behavior.
As expected, in both cases there is a certain extension of stability for the circular orbits around $r_U$, as discussed in subsection \ref{subsubsec:stability}. Finally, when $E^2>E_U^2$, the test particles that approach from a radial distance $r_i$, do not encounter any turning points and hence, have no other choice but to fall onto the event horizon. This way, the capture zone of the black hole is identified. In Fig. \ref{fig:capture-sim}, some examples of captured trajectories by the SDdS black hole have been demonstrated. 
As it can be observed, the closer the energy of the particles is to $E_U$, the more they tend to spiral orbits before being captured by the black hole. In this sense, at energy values close to $E_U$, the particles form an unstable circular orbit around $r_U$ and then fall inexorably onto the event horizon. 

In this section, we gave a full study to the motion of particles with non-zero initial angular momentum and their possible types of orbits. Accordingly, the particles could either deflect into or out of the black hole, or perform circular orbits with limited stability. Since all kinds of possible particle orbits have been studied so far, we close our discussion at this point and summarize our results in the next section.

\section{Summary and conclusions}\label{sec:conclusion}

In this work, we have focused on the exact analytic solutions of the equations of motion for massive particles moving in the exterior geometry of a four-dimensional SD black hole associated with a positive cosmological constant. In particular, we studied the exact analytic solutions of the equations of motion that arise for the radial and angular motion around an SDdS black hole with small quantum corrections. We first explored the causal structure of the spacetime and identified its event and cosmological horizons. We then pursued a canonical Lagrangian dynamics method to obtain the first-order differential equations of motion. For the case of radial motion, we classified the types of possible orbits in the context of the radial effective potential, and calculated the exact solutions, separately, for each of the cases. We showed that for the frontal radial scatterings, the equations of motion for proper and coordinate time result in degenerate hyper-elliptic integrals, to which, we gave exact analytical solutions in terms of 2-variable and 5-variable indefinite Lauricella hypergeometric functions. We then applied these solutions to simulate the radial orbits for the FSFK and FSSK. We showed that despite the fact that the comoving observers experience crossing the cosmological and event horizons within a finite amount of time, to the distant observers it takes infinite time for the test particles to pass the horizons. For the case of critical radial orbits, we presented the analytical solutions in terms of hyperbolic functions and showed that the test particles experience two distinct fates, and starting from a critical radius, they either escape to the cosmological horizon or are captured by the black hole. The same scenario holds for the comoving and distant observers. Switching to the study of angular orbits, we argued that the effective potential could offer only certain types of orbits for particles with non-zero initial angular momentum. Since the potential has no minimum, no planetary bound orbits are offered by the black hole. {It is, however, important to note that, for the vanishing running parameter, which corresponds to the Schwarzschild-de Sitter black hole, the effective potential acquires a minimum, so that the planetary bound orbits are also possible. Such cases have been studied extensively, for example in Refs.~\cite{cruz_geodesic_2005,hackmann_complete_2008,hackmann_geodesic_2008,olivares_motion_2011}. Furthermore, in Ref.~\cite{fathi_study_2022}, the motion of particles in the exterior geometry of a black hole with a linear quintessential term and cloud of strings has been investigated, where the spacetime metric can mimic the line element \eqref{eq:lapse1} with a vanishing cosmological constant. In particular, the term $(3M-r)\epsilon$ acts similar to the combination of cloud of strings and linear quintessence, because of which, the black hole can offer planetary bound orbits. However, as we demonstrated in the previous sections, such a possibility is eluded from the SDdS  black hole, for which both the linear and quadratic terms are available in the spacetime metric.} On the other hand, the potential's extremum defines a radius, at which, the particles can be on circular orbits with some extent of stability. We calculated the energy and angular momentum of the test particles on such orbits and demonstrated their radial profiles. We also inferred that the circular orbits at the vicinity of the potential's maximum can be stable since the second derivative of the characteristic polynomial is positive in a certain domain. We identified this domain by calculating the epicyclic frequencies of particles on circular orbits, around the potential's maximum. We also paid attention to the scattered trajectories which correspond to particles moving on the OFK. Such orbits occur when the initial energy of the particles is less than that at the potential's maximum, and hence, they can escape from the black hole. We showed that in general, the equation of motion for angular trajectories leads to a hyper-elliptic integral. First, we solved this equation to obtain the radial profile of the azimuth angle. The solution was given in terms of a 4-parameter Lauricella hypergeometric function and was then exploited to calculate the deflection angle of scattered particles. We plotted the changes of this angle in terms of the variations in the test particles' energy and showed that, as expected, it diverges at the vicinity of the energy of circular orbits. To study the behavior of particles on angular geodesics, we then performed an analytical treatment of the equation of motion, which involves the inversion of the included hyper-elliptic integral. This was a particular case of the Jacobi inversion problem, and hence, the process of obtaining the solution involved the abelian modular functions of genus two. We calculated the holomorphic and meromorphic differentials which are indispensable in the identification of the period matrices associated with the algebraic curve on the homologous Riemann surface. Accordingly, the general solution for the angular motion was expressed by the Kleinian sigma functions, which are given in terms of the Riemannian theta function of genus two with two-dimensional vectorial characteristics. Based on this solution, the orbits were discussed and simulated in accordance with the classifications offered by the effective potential. We plotted several cases of the OFK for  different turning points, and as expected, the orbits shift from being repulsive to being attractive, by approaching the potential's extremum. We also plotted several cases of the OSK. This was followed by demonstrating the critical orbits, which are comprised of unstable circular orbits, that either escape from the black hole or fall onto the event horizon, and hence, these two orbits are the upper limits of the OFK and OSK. We finally paid attention to the capture zone, for which, the incident particles with higher energies fall inexorably onto the black hole. In this sense, the critical orbits form the lower boundary of the capture zone. Note that, since the SDdS black hole is incapable of forming an accretion disk, it cannot be regarded as a real astrophysical black hole. However, studies like the one performed in this paper may equip scientists with advanced mathematical tools which pave the way to do rigorous scrutinization of other SD alternatives to general relativistic spacetimes with more similarity to real astrophysical black hole geometries and put them into observational assessments.

\section*{Acknowledgements}  
The author acknowledges Universidad de Santiago de Chile for financial support through the Proyecto POSTDOC-DICYT, C\'{o}digo 042331CM$\_$Postdoc. I would like to thank \'{A}ngel Rinc\'{o}n for introducing Ref. \cite{Panotopoulos:2021} and the SDdS solution.

\appendix

\section{Derivation of the radial solution of the FSFK}\label{app:A}

Applying the change of variable $r\rightarrow x/d_s$ to the equation of motion \eqref{eq:dotr_rad}, results in the equation
\begin{equation}
\tau(x) = -\frac{d_s}{\sqrt{\tL}}\int_1^{x}\frac{\ed x}{\sqrt{x^2(1-x)(d_s-d_f x)(d_s-d_1x)}},
    \label{eq:A1}
\end{equation}
which contains a degenerate hyper-elliptic integral. A second change of variable $x\rightarrow 1-z$, yields
\begin{equation}
\tau(z)=\frac{d_s}{\sqrt{\tL}}\int_0^{z}\frac{\ed z}{\sqrt{p_5(z)}},
    \label{eq:A2}
\end{equation}
in which
\begin{eqnarray}
p_5(z) &=& (1-z)^2 z (d_s-d_f[1-z])(d_s-d_1[1-z])\nonumber\\
&=& z (1-z)^2\left[1+\frac{d_f z}{d_s-d_f}\right](d_s-d_f)\left[1+\frac{d_1 z}{d_s-d_1}\right](d_s-d_1)\nonumber\\
&=&\ell^2 z (1-z)^2(1-c_1 z)(1-c_2 z).
\label{eq:A3}
\end{eqnarray}
This helps us recasting Eq.~\eqref{eq:A2} as
\begin{equation}
    \tau(z) = \frac{d_s}{\sqrt{\tL\ell^2}}\int_0^{z} z^{-\frac{1}{2}}(1-z)^{-1}(1-c_1 z)^{-\frac{1}{2}}(1-c_2 z)^{-\frac{1}{2}}\,\ed z.
    \label{eq:A4}
\end{equation}
Comparing the above relation to the one-dimensional integral form \cite{Akerblom:2005}
\begin{equation}
\int_0^{z} z^{a-1}(1-z)^{c-a-1}\prod_{i=1}^{n}(1-\xi_i z)^{-b_i}\,\ed z = 
\frac{z^{a}}{a}F_D^{(n+1)}\left(
a,b_1,\dots,b_n,1+a-c;a+1;\xi_1,\dots,\xi_n,z
\right)
    \label{eq:A5}
\end{equation}
of the incomplete $n$-variable Lauricella hypergeometric function, provides $n=2$, $a=c=b_1=b_2=1/2$, $\xi_1=c_1$ and $\xi_2=c_2$.

\section{{Expressing some of the solutions in terms of elliptic integrals}}\label{app:B}

{Applying the change of variable $r\rightarrow1/u$, and after some manipulations,  the differential equation \eqref{eq:dotr_rad} takes the form
\begin{equation}
\tau(u) = \frac{1}{\sqrt{\psi_0\tL}}\int_{u}^{u_s}\frac{\ed u}{u\sqrt{\left(u_f-u\right)\left(u_s-u\right)\left(u-u_1\right)}},
    \label{eq:B1}
\end{equation}
where $\psi_0=-d_sd_fd_1$, $u_s=1/d_s$, $u_f = 1/d_f$, and $u_1=1/d_1$, which respect the hierarchy $u_1<u<u_s<u_f$. Based on this condition, this integral can be re-expressed as \cite{byrd_handbook_1971}
\begin{equation}
\tau(u) = \frac{1}{\sqrt{\psi_0\tL}}\frac{\varrho}{u_s}\int_0^{y_1}\frac{\mathrm{dn}^2 y}{\left(1-\tilde{\alpha}\,\mathrm{sn}^2y\right)}\ed y
    \label{eq:B2}
\end{equation}
in which $\mathrm{sn}\,y\equiv\mathrm{sn}(y,\mathfrak{k})$ and $\mathrm{dn}\,y\equiv\mathrm{dn}(y,\mathfrak{k})$ are, respectively, the Jacobi elliptic sine function and the Jacobi delta amplitude with the modulus
\begin{equation}
\mathfrak{k}^2 =\frac{u_s-u_1}{u_f-u_1}, 
    \label{eq:B4}
\end{equation}
and the variable $y$ is defined in terms of the relation
\begin{equation}
\mathrm{sn}^2y=\frac{\left(u_f-u_1\right)\left(u_s-u\right)}{\left(u_s-u_1\right)\left(u_f-u\right)}.
    \label{eq:B5}
\end{equation}
This way, the limit of the upper integral \eqref{eq:B2} is given by $\mathrm{sn}\,y_1 = \sin\varphi$, where
\begin{equation}
\varphi = \mathrm{am}\,y_1 = \arcsin\left(\mathrm{sn}\,y\right),
    \label{eq:B6}
\end{equation}
is the Jacobi amplitude of the functions. Furthermore, we have notated
\begin{subequations}
\begin{align}
    & \varrho = \frac{2}{\sqrt{u_f-u_1}},\\
    & \tilde{\alpha} = \frac{\mathfrak{K}^2 u_f}{u_s}.
\end{align}
\label{eq:B7}
\end{subequations}
This way, the solution to the integral \eqref{eq:B2} can be expressed as \cite{byrd_handbook_1971}
\begin{equation}
\tau(u) = \frac{\varrho}{u_s\sqrt{\psi_0\tL}}\frac{ \mathfrak{k}^2}{\tilde{\alpha}^2}\sum_{j=0}^{1}\frac{\left(\tilde{\alpha}^2-\mathfrak{k}^2\right)^j}{\mathfrak{k}^{2j}j!\left(1-j\right)!}\mathcal{V}_j,
    \label{eq:B8}
\end{equation}
in which
\begin{subequations}
\begin{align}
& \mathcal{V}_0 = F\left(\varphi,\mathfrak{k}\right),\\
& \mathcal{V}_1 = \Pi\left(\varphi,\tilde{\alpha}^2,\mathfrak{k}\right),\\
\end{align}
\label{eq:B9}
\end{subequations}
are, respectively, the incomplete elliptic integrals of the first and third kind. The same procedure can be pursued for the differential equation \eqref{eq:drdt_rad}, which by means of the change of variable $r\rightarrow 1/u$ and partial fraction decomposition, can be recast as
\begin{multline}
t(u)=\frac{E}{\sqrt{\psi_0\tL^3}}\Bigg[-\frac{r_{++}}{\psi_1}\int_{u}^{u_s}\frac{\ed u}{\left(u_{++}-u\right)\sqrt{\left(u_f-u\right)\left(u_s-u\right)\left(u-u_1\right)}}\\
+\frac{r_{+}}{\psi_2}\int_{u}^{u_s}\frac{\ed u}{\left(u_{+}-u\right)\sqrt{\left(u_f-u\right)\left(u_s-u\right)\left(u-u_1\right)}}\\
+\frac{r_{3}}{\psi_3}\int_{u}^{u_s}\frac{\ed u}{\left(u_{3}-u\right)\sqrt{\left(u_f-u\right)\left(u_s-u\right)\left(u-u_1\right)}}\Bigg],
\label{eq:B10}    
\end{multline}
with $u_{++}=1/r_{++}$, $u_+=1/r_+$ and $u_3=1/r_3$, and by defining $\psi_1=r_{++}(r_{++}-r_+)(r_{++}-r_3)$, $\psi_2=r_+(r_{++}-r_+)(r_+-r_3)$ and $\psi_3=-r_3(r_{++}-r_3)(r_+-r_3)$. The solution to this equation is given by \cite{byrd_handbook_1971}
\begin{multline}
t(u)=\frac{E}{\sqrt{\psi_0\tL^3}}\Bigg[\frac{r_{++}\varrho}{\psi_1\left(u_{s}-u_{++}\right)}\frac{\mathfrak{k}^2}{\tilde{\alpha}_{++}^2}\sum_{j=0}^{1}\frac{\left(
\tilde{\alpha}_{++}^2-\mathfrak{k}^2
\right)^j}{\mathfrak{k}^{2j}j!\left(1-j\right)!}\mathcal{V}_{++j}\\
+\frac{r_{+}\varrho}{\psi_2\left(u_{+}-u_s\right)}\frac{\mathfrak{k}^2}{\tilde{\alpha}_{+}^2}\sum_{j=0}^{1}\frac{\left(
\tilde{\alpha}_{+}^2-\mathfrak{k}^2
\right)^j}{\mathfrak{k}^{2j}j!\left(1-j\right)!}\mathcal{V}_{+j}\\
-\frac{r_{3}\varrho}{\psi_3\left(u_{s}-u_3\right)}\frac{\mathfrak{k}^2}{\tilde{\alpha}_{3}^2}\sum_{j=0}^{1}\frac{\left(
\tilde{\alpha}_{3}^2-\mathfrak{k}^2
\right)^j}{\mathfrak{k}^{2j}j!\left(1-j\right)!}\mathcal{V}_{3j}\Bigg],
    \label{eq:B11}
\end{multline}
in which $\mathcal{V}_{++j}$, $\mathcal{V}_{+j}$ and $\mathcal{V}_{3j}$ have the same expressions as in Eqs. \eqref{eq:B9}, considering the respected exchanges $\tilde{\alpha}\rightarrow\tilde{\alpha}_{++}, \tilde{\alpha}_+,\tilde{\alpha}_3$, where
\begin{subequations}
\begin{align}
    & \tilde{\alpha}_{++}^2 = \frac{\mathfrak{k}^2\left(u_f-u_{++}\right)}{\left(u_s-u_{++}\right)},\\
    & \tilde{\alpha}_{+}^2 = \frac{\mathfrak{k}^2\left(u_{+}-u_f\right)}{\left(u_{+}-u_s\right)},\\
     & \tilde{\alpha}_{3}^2 = \frac{\mathfrak{k}^2\left(u_{f}-u_3\right)}{\left(u_{s}-u_3\right)}.
     \end{align}
\label{eq:B12}
\end{subequations}
The integral in Eq. \eqref{eq:phi(r)-int} is genuinely hyper-elliptic, and we note that the only way to express the solutions in terms of ordinary elliptic integrals is on one of the limits $r\gg r_F$, $|r_4|\ll 1$, or $r\gg|r_5|$. Under these conditions, the solution of the integral \eqref{eq:phi(r)-int} can be obtained in a similar way as in Eq. \eqref{eq:B8}. 
}

\bibliographystyle{ieeetr}
\bibliography{biblio.bib}

\end{document}